
\input harvmac

\font\zfont = cmss10 
\font\litfont = cmr6

\def\bigone{\hbox{1\kern -.23em {\rm l}}}
\def\ZZ{\hbox{\zfont Z\kern-.4emZ}}
\def\half{{\litfont {1 \over 2}}}

\def\rank{{\rm rank ~}}

\noblackbox

\def\np#1#2#3{Nucl. Phys. B{#1} (#2) #3}
\def\pl#1#2#3{Phys. Lett. {#1}B (#2) #3}

\def\physrev#1#2#3{Phys. Rev. {D#1} (#2) #3}

\def\prep#1#2#3{Phys. Rep. {#1} (#2) #3}

\def\tilde{\widetilde}

\def\Tr{{\rm Tr ~}}
\def\ev#1{\langle#1\rangle}

\def\CW{{\cal W}}
\def\tilde{\widetilde}
\def\half{{\litfont {1 \over 2}}}

\def\CC{{\cal C}}

\Title{hep-th/9503179, RU-95-3, IASSNS-HEP-95/5}
{\vbox{\centerline{Duality, Monopoles, Dyons, Confinement}
\centerline{and Oblique Confinement in}
\centerline{Supersymmetric $SO(N_c)$ Gauge Theories}}}
\bigskip
\centerline{K. Intriligator$^1$ and N. Seiberg$^{1,2}$}
\vglue .5cm
\centerline{$^1$Department of Physics and Astronomy}
\centerline{Rutgers University}
\centerline{Piscataway, NJ 08855-0849, USA}
\vglue .3cm
\centerline{$^2$ School of Natural Sciences}
\centerline{Institute for Advanced Study}
\centerline{Princeton, NJ 08540, USA}

\bigskip

\noindent
We study supersymmetric $SO(N_c)$ gauge theories with $N_f$ flavors of
quarks in the vector representation.  Among the phenomena we find are
dynamically generated superpotentials with physically inequivalent
branches, smooth moduli spaces of vacua, confinement and oblique
confinement, confinement without chiral symmetry breaking, massless
composites (glueballs, exotics, monopoles and dyons), non-trivial fixed
points of the renormalization group and massless magnetic quarks and
gluons.  Our analysis sheds new light on a recently found duality in
$N=1$ supersymmetric theories.  The dual forms of some of the theories
exhibit ``quantum symmetries'' which involve non-local transformations
on the fields.  We find that in some cases the duality has both $S$ and
$T$ transformations generating $SL(2,Z)$ (only an $S_3$ quotient of which
is realized non-trivially).  They map the original non-Abelian electric
theory to magnetic and dyonic non-Abelian theories.  The magnetic
theory gives a weak coupling description of confinement while the dyonic
theory gives a weak coupling description of oblique confinement.  Our
analysis also shows that the duality in $N=1$ is a generalization of the
Montonen-Olive duality of $N=4$ theories.

\Date{3/95}

\newsec{Introduction}
\lref\sv{M.A. Shifman and A. I. Vainshtein, \np{277}{1986}{456};
\np{359}{1991}{571}}
\nref\nonren{N. Seiberg, hep-ph/9309335, \pl{318}{1993}{469}}%
\nref\nati{N. Seiberg, hep-th/9402044, \physrev{49}{1994}{6857}}%
\nref\kl{V. Kaplunovsky and J. Louis, hep-th/9402005,
\np{422}{1994}{57}}%
\nref\ils{K. Intriligator, R.G. Leigh and N. Seiberg, hep-th/9403198,
\physrev{50}{1994}{1092}; K. Intriligator, hep-th/9407106,
\pl{336}{1994}{409}}%
\nref\swi{N. Seiberg and E. Witten, hep-th/9407087, \np{426}{1994}{19}}%
\nref\swii{N. Seiberg and E. Witten, hep-th/9408099,
\np{431}{1994}{484}}%
\nref\intse{K. Intriligator and N. Seiberg, hep-th/9408155,
\np{431}{1994}{551}}%
\nref\sunnt{A. Klemm, W. Lerche, S. Theisen and S. Yankielowicz,
hep-th/9411048, hep-th/9412158; P. Argyres and A. Faraggi, hep-th/9411057}
\nref\iss{K. Intriligator, N. Seiberg and S. Shenker, hep-ph/9410203,
\pl {342}{1995}{152}}%
\nref\sem{N. Seiberg, hep-th/9411149 , RU-94-82, IASSNS-HEP-94/98}%
\nref\aharony{O. Aharony, hep-th/9502013, TAUP-2232-95}%
\nref\kutasov{D. Kutasov, hep-th/9503086, EFI-95-11}%
\nref\rlms{R. Leigh and M. Strassler, RU-95-2, hep-th/9503121}%
\nref\doush{M. Douglas and S. Shenker, RU-95-12, to appear}%
\nref\finnpou{D. Finnell and P. Pouliot, RU-95-14, SLAC-PUB-95-6768,
hep-th/9503115}%

Recently, it has become clear that certain aspects of four dimensional
supersymmetric field theories can be analyzed exactly, thus providing a
laboratory for the analysis of the dynamics of gauge theories
\refs{\nonren-\finnpou} (for a recent short review, see
\ref\pwer{N. Seiberg, The Power of Holomorphy -- Exact Results in 4D
SUSY Field Theories.  To appear in the Proc. of PASCOS 94.
hep-th/9408013, RU-94-64, IASSNS-HEP-94/57}).
\nref\ads{I. Affleck, M. Dine and N. Seiberg, \np{241}{1984}{493};
\np{256}{1985}{557}}%
\nref\nsvz{V.A. Novikov, M.A. Shifman, A. I.  Vainshtein and V. I.
Zakharov, \np{223}{1983}{445}; \np{260}{1985}{157}}%
\nref\rusano{V. Novikov, M. Shifman, A. Vainshtein and V.
Zakharov, \np{229}{1983}{381}}%
\nref\cern{D. Amati, K. Konishi, Y. Meurice, G.C. Rossi and G.
Veneziano, \prep{162}{1988}{169} and references therein}%
Most of the work so far was devoted to $SU(N_c)$ gauge theories (see
\refs{\ads-\cern} for earlier work on these theories), which exhibited
interesting physical phenomena.  It is natural to ask which of these
results are specific to $SU(N_c)$ and which are more general.
Furthermore, other theories might exhibit qualitatively new phenomena.
Here we address these questions by studying $SO(N_c)$ gauge theories
with $N_f$ flavors of quarks, $Q^i$ ($i=1,...,N_f$), in the
vector representation of the gauge group.
Another motivation to study these theories is associated with the role
of the center of the group in confinement.  Unlike the $SU(N_c)$
theories with quarks in the fundamental representation, where there is
no invariant distinction between Higgs and confinement
\ref\higgscon{T. Banks, E. Rabinovici, \np{160}{1979}{349};
E. Fradkin and S. Shenker, \physrev{19}{1979}{3682}.},
here these phases (as well as the oblique confinement
phase\nref\thooft{G. 'tHooft, \np{190}{1981}{455}}
\nref\cardyrabin{J. Cardy and E. Rabinovici,
\np{205}{1982}{1};
J. Cardy, \np{205}{1982}{17}}\refs{\thooft, \cardyrabin})
can be distinguished.  A Wilson loop in
the spinor representation cannot be screened by the dynamical quarks and
therefore it is a gauge invariant order parameter for confinement.  A
similar comment applies to the 'tHooft loop and to the product of the
'tHooft loop and the Wilson loop (which probes oblique confinement).

Our results may be summarized as follows:

\medskip

\noindent
Theories with $N_f<N_c-4$ have dynamically generated superpotentials,
associated with gaugino condensation, similar to the ones found in
$SU(N_c)$ theories with $N_f < N_c$ \ads.

\medskip

\noindent
Theories with $N_f=N_c-4$ have two physically inequivalent phase
branches.  One branch has a dynamically generated superpotential, just
as with $N_f<N_c-4$.  On the other branch no superpotential is
generated; on this branch there is a moduli space of physically
inequivalent but degenerate quantum vacua.  This moduli space of vacua
differs from the classical one in that a classical singularity at the
origin is smoothed out in the quantum theory in a fashion similar to
$SU(N_c)$ gauge theories with $N_f=N_c$ \nati\ and to an $SU(2)$ gauge
theory with quarks in the $I=3/2$ representation \iss.
The theory at the origin of this space has
confinement without chiral symmetry breaking.

\medskip

\noindent
For $N_f=N_c-3$, we again find two physically inequivalent phase
branches, one with a dynamically generated superpotential and the other
with a moduli space of quantum vacua.  In the branch with the moduli
space of vacua there is again confinement without chiral symmetry
breaking and there are $N_f$ massless composite fields at the origin.
They can be interpreted as glueballs (for $N_c=4$) or exotics (for $N_c
>4$).  This phenomenon of massless composites is similar to the one
found in $SU(N_c)$ theories with $N_f=N_c+1$ \nati.

\medskip

\noindent
Theories with $N_f=N_c-2$ have no superpotential -- there is, again, a
quantum moduli space of degenerate vacua.  The low energy theory
contains a massless photon and hence the theory is in a Coulomb phase.
As in \refs{\swi-\sunnt} we exactly compute its effective gauge
coupling on the quantum moduli space of vacua.  We find various
numbers of massless monopoles and dyons at various vacua at strong
$SO(N_c)$ coupling.  As in \swii, they transform non-trivially under
the flavor symmetry.  When a mass term is added the monopoles and
dyons condense leading to confinement and oblique confinement
\refs{\thooft,\cardyrabin} respectively.  These correspond to the
two physically inequivalent branches of the theories with $N_f=N_c-3,
N_c-4$.

\medskip

\noindent
For $N_f> N_c-2$, the theory at the origin of the space of vacua is
in a non-Abelian Coulomb phase.  It can be given a dual ``magnetic''
description in terms of an $SO(N_f-N_c+4)$ theory with matter discussed
in \sem.  The dynamical scale $\tilde \Lambda$ of the dual theory is
inversely related to the scale $\Lambda$ of the original theory
\eqn\ltlmu{\Lambda ^{3(N_c-2)-N_f}\tilde \Lambda
^{2N_f-3(N_c-2)} \sim \mu ^{N_f}}
and therefore the electric theory becomes weaker as the magnetic theory
becomes stronger and vice versa.  The meaning of this relation and of
the scale $\mu$ will be explained in detail.  The interpretation of the
dual theory as being ``magnetic'' becomes obvious in the special case
$N_f=N_c-2$ where the low energy gauge group is $U(1)$ and the dual
matter fields are the magnetic monopoles.  The duality for $N_f>N_c-2$
is then clearly identified as a non-Abelian generalization of ordinary
electric-magnetic duality.  For $N_c-2< N_f \le {3 \over 2} (N_c-2)$ the
magnetic degrees of freedom are free in the infra-red while for ${3
\over 2} (N_c-2)< N_f < 3 (N_c-2)$ the electric and the magnetic
theories flow to the same non-trivial fixed point of the renormalization
group.

In sections 2 -- 4 we discuss all these cases.  We start from a small
number of flavors and gradually consider larger $N_f$.  We then check
that our results fit together upon giving the quarks $Q^i$ masses and
reducing the number of flavors.

One lesson from these theories is that the qualitative phenomena found
in $SU(N_c)$ theories and some $N=2$ theories are more generic and apply
in a wider class of $N=1$ theories. Other lessons are associated with
the new subtleties which are specific to these theories.

In section 5 we discuss $SO(3)$ gauge theories with $N_f$ quarks.  They
exhibit new complications which are not present for larger values of
$N_c$.  Some of their dual theories exhibit quantum symmetries.  These
are symmetries which are not easily visible from the Lagrangian because
they are implemented by non-local transformations on the fields.

The $SO(3)$ theories lead us to the first example of new duality
transformations in $N=1$ theories.  The electric theory can be
transformed both to a magnetic and to a dyonic theory.  The electric
theory is weakly coupled in the Higgs branch of the theory (along the
flat directions) and strongly coupled in the confining and the oblique
confining branches of the theory.  The magnetic (dyonic) theory is
weakly coupled in the confining (oblique confinement) branch of the
theory and is strongly coupled in the Higgs and the oblique confinement
(confinement) branches.  The confining and the oblique confinement
branches of the theory are related by a spontaneously broken global
discrete symmetry.  Therefore, the magnetic and the dyonic theories are
similar.  The group of duality transformations which permutes these
theories is $S_3$.  It is related to the standard duality group
$SL(2,Z)$ by a quotient by $\Gamma(2)$, which acts trivially on the
theories.  In other words, the duality transformation $S \in SL(2,Z) $
relates the electric theory to the magnetic theory while $T \in SL(2,Z)$
maps the magnetic theory to the dyonic theory.

The analysis of $SO(3)$ with $N_f=3$ establishes the relation between
the Montonen-Olive duality
\ref\om{C. Montonen and D. Olive, \pl {72}{1977}{117}; P. Goddard,
J. Nuyts and D. Olive, \np{125}{1977}{1}}
of $N=4$ theories
\ref\dualnf{H. Osborn, \pl{83}{1979}{321}; A. Sen, hep-th/9402032,
\pl{329}{1994}{217}; C. Vafa and E. Witten, hep-th/9408074,
\np{432}{1994}{3}}
and the duality in $N=1$ theories \sem.  When a generic cubic
superpotential is turned on the theory flows in the infra-red to an
$N=4$ theory.  Its $N=1$ dual is an $SO(4)$ theory, with $N_f=3$, which
flows in the infra-red to an $SU(2)$ $N=4$ theory\foot{Throughout this
paper we limit ourselves to the Lie algebra and do not discuss the role
of the center of the gauge group.  Therefore we do not distinguish
between $SO(N_c)$ and $Spin(N_c)$.}.  These two similar $N=4$ theories
are dual to each other as in \refs{\om, \dualnf}.  Therefore, the
duality in $N=1$ theories \sem\ is compatible with and generalizes the
Montonen-Olive duality of $N=4$.

In section 6 we present more dyonic theories for $N_f=N_c-1$.  Unlike
the dyonic theories discussed in section 5, here there is no global
symmetry which makes the theories similar.  The electric, magnetic and
dyonic theories are really distinct.  As in section 5, the electric
theory gives a weak coupling description of the Higgs branch of the
theory, the magnetic theory gives a weak coupling description of the
confining branch and the dyonic theory gives a weak coupling description
of the oblique confinement branch of the theory.  In all these examples
the magnetic theory is weakly coupled at the non-Abelian Coulomb point
while the other two theories are strongly coupled there.

\newsec{Preliminaries}

\subsec{The classical moduli spaces for $SO(N_c)$ with $N_f$ quarks
$Q^i$.}

The classical theory with $N_f$ massless quarks has a moduli space of
degenerate vacua labeled by the expectation values $\ev {Q}$ of the
scalar components of the matter fields subject to the ``D-flatness''
constraints.  Up to gauge and global rotations, the solutions of these
equations for $N_f< N_c$ are of the form
\eqn\flcqev{Q=\pmatrix{a_1&\ &\ &\ &\ &\ \cr\ &a_2&\ &\ &\ &\
\cr\ &\ &\ . &\ &\ &\ \cr\ &\ &\ &a_{N_f}&\ &\ \cr}}
where, using gauge transformations, the sign of any $a_i$ can be
flipped.  For generic $a_i$ these expectation values break $SO(N_c)$ to
$SO(N_c-N_f)$ by the Higgs mechanism for $N_f \le N_c - 2$ and
completely break $SO(N_c)$ for $N_f > N_c-2$.
For $N_f\geq N_c$ the flat directions are of the form
\eqn\fmcqev{Q=\pmatrix{a_1&\ &\ &\ \cr\ &a_2&\ &\ \cr\ &\ &.&\ \cr\ &\
&\ &a_{N_c}\cr\ &\ &\ &\ \cr\ &\ &\ &\ \cr}.}
If some of the $a_i$ vanish, the signs of the others can be flipped by
gauge transformations.  However, if all $a_i\neq 0$, gauge
transformations can only be used to flip signs in such a way that the
product $\prod _{i=1}^{N_c}a_i$ is invariant.

For $N_f<N_c$ the space of vacua can be given a gauge invariant
description in terms of the expectation values of the ``meson'' fields
$M^{ij}=Q^i\cdot Q^j$.  These $\half N_f(N_f+1)$ superfields correspond
precisely to the matter superfields left massless after the Higgs
mechanism.  Their expectation values are classically unconstrained.  For
$N_f\geq N_c$ it is also possible to form ``baryons'' $B^{[i_1\dots
i_{N_c}]}=(Q)^{N_c}$, with indices antisymmetrized.  Along the flat
direction \fmcqev\ we have
\eqn\mbfd{\eqalign{M&=\pmatrix{a_1^2&\ &\ &\ &\ &
\cr\ &a_2^2&\ &\ &\ &\
\cr\ &\ &.&\ &\ &\
\cr\ &\ &\ &a_{N_c}^2&\ &\
\cr\ &\ &\ &\ &\ &\
\cr\ &\
&\ &\ &\ &\ \cr}
\cr B^{1,\dots , N_c}&=a_1\dots a_{N_c}}}
with all other components of $M$ and $B$ vanishing.  $M$ thus has rank
at most $N_c$.  If the rank of $M$ is less than $N_c$, $B=0$.  If the
rank of $M$ is $N_c$, $B$ has rank one and its non-zero eigenvalue is,
with an undetermined sign, the square root of the product of non-zero
eigenvalues of $M$.  The classical moduli space of vacua for $N_f\geq
N_c$ is therefore described by the space of $M$ of rank at most $N_c$
along with a sign, corresponding to the sign of $B=\pm \sqrt{\det 'M}$,
for $M$ of rank $N_c$.

\subsec{The quantum theories and some conventions}

The quantum $SO(N_c)$ theory with $N_f$ massless quarks has an anomaly
free global $SU(N_f)\times U(1)_R$ symmetry with the fields $Q$
transforming as $(N_f)_{N_f-N_c+2\over N_f}$.
In addition, the theory is invariant
under the $Z_2$ charge conjugation symmetry $\CC$ and the $Z_{2N_f}$
($Z_{4N_f}$ for $N_c=3$) discrete symmetry
\eqn\disc{\eqalign{Q&\rightarrow e^{2\pi i/2N_f}Q\cr Q&\rightarrow
e^{2\pi i/4N_f}Q\qquad \hbox{for}\qquad N_c=3}}
(its $Z_{N_f}$ subgroup is also in the center of $SU(N_f)$).

For $N_c>4$ the one-loop beta function implies that the gauge coupling
runs as $e^{-8\pi ^2g^{-2}(E)+i\theta}=(\Lambda _{N_c,N_f}/E)^{3(N_c-2)-N_f}$,
where $\Lambda _{N_c, N_f}$ is the dynamically
generated scale for the theory with $N_f$ quarks.  By adding the
$\theta$ angle, the scale $\Lambda$ becomes a complex number, which can
be interpreted as the first component of a chiral superfield.
Since $SO(4)\cong
SU(2)_L\times SU(2)_R$, there are independent running gauge couplings
for each $SU(2)_s$:  $e^{-8\pi ^2g_s^{-2}(E)+i\theta}=
(\Lambda _{s,N_f}/E)^{6-N_f}$
for $s=L,R$.  For $N_c=3$ the gauge coupling runs
as $e^{-8\pi ^2g^{-2}(E)+i\theta}=(\Lambda _{3,N_f}/E)^{6-2N_f}$.

By giving the quarks $Q^{N_f}$ a mass term with $W_{tree}= \half
mM^{N_fN_f}$ and decoupling the massive matter, the theory with $N_f$ quarks
yields a low energy theory with $N_f-1$ quarks.   Matching the running gauge
coupling at the mass scales where the massive quarks decouple relates
the scale of the original high-energy theory to
the scale of the low energy Yang-Mills theory as
\eqn\massmatch{\eqalign{\Lambda _{N_c, N_f}^{3(N_c-2)-N_f}m&=\Lambda
_{N_c,N_f-1}^{3(N_c-2)-(N_f-1)}\cr
\Lambda _{s,N_f}^{6-N_f}m&=\Lambda _{s,N_f-1}^{6-(N_f-1)}\cr
 \Lambda _{3,N_f}^{6-2N_f}m^2&=\Lambda
_{3,N_f-1}^{6-2(N_f-1)}}\qquad\eqalign{\hbox{for}\ N_c&>4\cr
\hbox{for}\ N_c&=4,\ \ s=L,R\cr \hbox{for}\ N_c&=3\cr}.}
The absence of any constant ``threshold'' factors in these matching
relations define our normalization for $\Lambda _{N_c,N_f}$ relative
to the normalization of $\Lambda _{N_c,0}$.  (For $N_c=3$ this
convention differs slightly from the one used in \intse.)  For more
discussion on the threshold factors in these theories see \finnpou.

The $SO(N_c)$ theory with $N_f$ quarks can also be related to an
$SO(N_c-1)$ theory with $N_f-1$ quarks via the Higgs mechanism
by taking the expectation
value $a_{N_f}$ in \flcqev\ to be large.  The scale $\Lambda _{N_c-1,
N_f-1}$ of the low energy theory is related to the scale $\Lambda
_{N_c,N_f}$ of the original theory by matching the running gauge coupling
at the energy scale set by $a_{N_f}$,
\eqn\higgsmatch{\eqalign{\Lambda _{N_c,
N_f}^{3(N_c-2)-N_f}(M^{N_fN_f})^{-1}&=\Lambda
_{N_c-1,N_f-1}^{3(N_c-2)-N_f-2}\cr
\Lambda _{5,N_f}^{9-N_f}(M^{N_fN_f})^{-1}&=\Lambda
_{s,N_f-1}^{6-(N_f-1)}\cr
4\Lambda _{L,N_f}^{6-N_f}\Lambda _{R,N_f}^{6-N_f}
(M^{N_fN_f})^{-2}&=\Lambda
_{3,N_f-1}^{6-2(N_f-1)}.}\qquad \eqalign{\hbox{for}\ N_c&>5\cr
\hbox{for}\ s&=L,R\cr &}}
For $N_c=4$, the quarks are in the $(2,2)$ representation of
$SU(2)_L\times SU(2)_R$ and we have $M^{N_fN_f}=Q^{N_f}\cdot Q^{N_f}=
Q^{N_f}_{c_Lc_R}Q^{N_f}_{d_Ld_R}\epsilon ^{c_Ld_L}\epsilon ^{c_Rd_R}$.
The factor of four in the last relation of \higgsmatch\ reflects the
fact that the natural order parameter in terms of the $SU(2)_s$ is
$\half M^{N_fN_f}$.  These relations define our relative
normalization of the $\Lambda _{N_c,N_f}$ for different $N_c$.  To fix
the absolute normalization, we use the conventions of \ils\ (see also
\finnpou).

With the scale normalizations defined above, gaugino condensation in
the pure $SO(N_c)$ Yang-mills theory is found to be given by
\eqn\gcgen{\eqalign{\ev{\lambda \lambda}&=\half 2^{4/(N_c-2)}\epsilon
_{N_c-2}\Lambda _{N_c,0}^3\cr
\ev{(\lambda \lambda)_s}&=\epsilon _s\Lambda _{s,0}^3\cr
\ev{\lambda \lambda}&=\epsilon _2\Lambda
_{3,0}^3}\qquad\eqalign{\hbox{for}\ N_c&\geq 5\cr \hbox{for}\ N_c&=4,\
s=L,R\cr \hbox{for}\ N_c&=3,}}
where $\epsilon _{N_c-2}$ is an $(N_c-2)$-th root of unity, reflecting
the $N_c-2$ (physically equivalent) supersymmetric vacua of the pure
gauge supersymmetric $SO(N_c)$ theory and, likewise, $\epsilon _s=\pm
1$ for $s=L,R$ and $\epsilon_2=\pm1$.  The last two equations follow
the convention for $SU(2)$ of \ils.  The first one can be derived by
studying the theory with matter and perturbing it with mass terms or
along the flat directions as in section 3 (see also
\finnpou).

Throughout most of this paper, we will limit the discussion of $SO(4)$
to the case $\Lambda_L=\Lambda_R$, which is similar to larger values of
$N_c$.

The classical vacuum degeneracy of \flcqev\ and \fmcqev\ can be lifted
by quantum effects.  In the low energy effective theory this is
represented by a dynamically generated superpotential for the light
meson fields $M^{ij}$.
Holomorphy and the $SU(N_f)\times U(1)_R$ symmetries
determine that any dynamically generated superpotential (for $N_c\neq
3$) must be of the form
\eqn\wsymm{W=A_{N_c,N_f}\big({\Lambda _{N_c,N_f}^{3N_c-N_f-6}\over
\det_{ij}( Q^i\cdot Q^j)}\big)^{1/(N_c-N_f-2)}=A_{N_c,N_f}\big({\Lambda
_{N_c,N_f}^{3N_c-N_f-6}\over \det M}\big)^{1/(N_c-N_f-2)},} for some
constants $A_{N_c,N_f}$.  For $N_f=N_c-2$ the superpotential \wsymm\
does not make sense.  For $N_c-2<N_f\leq N_c$, the superpotential
\wsymm\ cannot be generated because it has incorrect asymptotic
behavior in the limit of large $|Q|$, where asymptotic freedom implies
that any dynamically generated superpotential must be smaller than
$|Q|^3$.  For $N_f>N_c$, $\det Q^i\cdot Q^j=0$ and so the
superpotential \wsymm\ cannot exist.  In short, there can be no
dynamically generated superpotential for $N_f\geq N_c-2$.  Similarly,
(in the Higgs phase) there can be no dynamically generated
superpotential for $N_c=3$ for any $N_f$.  These theories have a
quantum moduli space of exactly degenerate but physically inequivalent
vacua.  We find that these theories display other interesting
non-perturbative gauge dynamics.  Even for $N_f < N_c-2$ where a
consistent invariant superpotential exists, we will show that it is
not always generated.

\newsec{The quantum theories for $N_c\geq 4$, $N_f \leq N_c-2$}

\subsec{$N_f\leq N_c-5$; a dynamically generated superpotential by
gaugino condensation.}

As in \ads, the superpotential \wsymm\ is generated by
gaugino condensation in the $SO(N_c-N_f)$ Yang-Mills theory
left unbroken by the $\ev{Q}$:
$W=(N_c-N_f-2)\ev{\lambda \lambda}$; the details
leading to the normalization factor were discussed in \ils\ for
$SU(N_c)$ gaugino condensation. Following the conventions discussed in
the previous section, this gives
\eqn\wi{W=\half (N_c-N_f-2)\epsilon _{(N_c-N_f-2)} \left({16\Lambda
_{N_c,N_f}^{3N_c-N_f-6}\over \det M}\right)^{1/(N_c-N_f-2)}.}
This quantum effective superpotential lifts the classical vacuum
degeneracy.  Indeed, the theory \wi\ has no vacuum at all.
Adding mass terms $W_{tree}=\half \Tr\ mM$ to the dynamically generated
superpotential \wi\ gives a theory with $(N_c-2)$ supersymmetric vacua:
\eqn\somev{\ev{M^{ij}}=\epsilon _{(N_c-2)} \left(16\det m\Lambda
_{N_c,N_f}^{3(N_c-2)-N_f}\right)^{1/(N_c-2)}\left( {1 \over m}
\right)^{ij}.}

If some of the masses are zero, we can integrate out the massive
quarks to find an effective superpotential for the massless ones.  It
is of the form \wi , with the scale of the low energy theory with fewer
quarks given by \massmatch.

\subsec{$N_f=N_c-4$; two inequivalent branches -- confinement without
chiral symmetry breaking}

In this case the $\ev{Q}$ break $SO(N_c)$ to $SO(4)\cong SU(2)_L\times
SU(2)_R$.  With the conventions discussed above,
the scales $\Lambda _{s,0}$ of the low energy
$SU(2)_s$ Yang-Mills theories are related to the scale of the high
energy theory by $\Lambda _{L,0}^6=\Lambda _{R,0}^6=\Lambda
^{2(N_c-1)}_{N_c, N_c-4}/\det M$.  Gaugino
condensation in the unbroken $SU(2)_L\times SU(2)_R$ generates the
superpotential
\eqn\wii{W=2\ev{\lambda \lambda }_L+2\ev {\lambda \lambda }_R=\half (\epsilon
_L+\epsilon _R)\left({16\Lambda ^{2(N_c-1)}_{N_c, N_c-4}\over \det
M}\right)^{1/2},}
where $\epsilon _L$ and $\epsilon _R$ are $\pm 1$ and the factors of two
follow from the discussion in \ils.

The $\epsilon _s$ in \wii\ reflect the fact that the low energy theory
has four ground states.  The two ground states with $\epsilon_L=
\epsilon _R$ are physically equivalent; they are related by a discrete R
symmetry.  The two ground states with $\epsilon _L=-\epsilon_R$ are
also physically equivalent.  However, the ground states with $\epsilon
_L=\epsilon _R$ are physically distinct from those with $\epsilon
_L=-\epsilon _R$.  The sign of $\epsilon _L\epsilon _R$ labels two
physically inequivalent phase branches of the low energy effective
theory.  The branch of \wii\ with $\epsilon _L\epsilon _R=1$ is simply
the continuation of \wi\ to $N_f=N_c-4$.  It lifts the classical vacuum
degeneracy and the quantum theory has no vacuum.

The two ground states with $\epsilon _L\epsilon _R=-1$ are different.
The superpotential \wii\ is then zero; there is a quantum moduli space
of degenerate but physically inequivalent vacua labeled by $\ev{M}$
(the two different values of $\epsilon_L$ on this branch mean that for
every $\ev{M}$ there are two ground states).  In particular, there is
a vacuum at the origin, $M=0$.  Classically, the low energy effective
theory has a singularity at the origin, corresponding to the
$SO(N_c)/SO(4)$ vector bosons which become massless there.  This
singularity shows up in the classical Kahler potential
$K_{classical}(M, M^\dagger)$.  In the quantum theory such a
singularity is either smoothed out or it is associated with some
fields which become massless there.  Our result is that the classical
singularity at the origin is simply smoothed out.  In other words, the
massless spectrum at the origin is the same as it is elsewhere,
consisting simply of the fields $M$.

This result satisfies several independent and highly non-trivial
consistency conditions.  For example, because the theory has a global
$SU(N_f)\times U(1)_R$ symmetry which is unbroken at the origin, this
assertion about the massless spectrum at the origin can be checked
using the 't Hooft anomaly matching conditions.  The classical
massless fermions are the quark components of the $Q^i$, with the
global quantum numbers $N_c\times (N_f)_{{2-N_c\over N_f}}$, and the
gluinos with the global quantum numbers $\half N_c(N_c-1)\times
(1)_{1}$. This classical massless spectrum gives for the 't Hooft
anomalies
\eqn\micro{\eqalign{U(1)_R\qquad &-\half N_c(N_c-3)\cr
U(1)_R^3\qquad &\half N_c(N_c-1)+{N_c\over N_f^2}(2-N_c)^3\cr
SU(N_f)^3\qquad &N_cd_3(N_f)\cr
SU(N_f)^2U(1)_R\qquad &N_c({2-N_c\over N_f})d_2(N_f),}}
where $d_2(N_f)$ and $d_3(N_f)$ are the quadratic and cubic $SU(N_f)$
Casimirs in the fundamental representation.  Our asserted massless
fermionic spectrum at the origin in the quantum theory is simply the
fermionic component of $M$, with the global quantum numbers $(\half
N_f(N_f+1))_{N_f-2N_c+4\over N_f}$;  this field has the 't Hooft
anomalies
\eqn\manom{\eqalign{U(1)_R\qquad &\half (N_f+1)(N_f-2N_c+4)\cr
U(1)_R^3\qquad &\half N_f^{-2}(N_f+1)(N_f-2N_c+4)^3\cr
SU(N_f)^3\qquad &(N_f+4)d_3(N_f)\cr
SU(N_f)^2U(1)_R\qquad &(N_f+2)({N_f-2N_c+4\over N_f})d_2(N_f).}}
It is non-trivial but true that these anomalies match the anomalies
\micro\  of the microscopic theory for $N_f=N_c-4$.   As in \iss, we use
this fact as evidence that the Kahler potential near the origin is
smoothed out by quantum effects: $K(M\rightarrow 0)\sim \Tr\ M^\dagger
M/|\Lambda |^2$.

Consider now giving $Q^{N_f}$ a mass and integrating it out.  The
resulting low energy theory should agree with our prior results for
$N_f=N_c-5$.  For the branch of \wii\ with $\epsilon _L\epsilon_R=1$,
adding $W_{tree}=\half mM^{N_fN_f}$ and integrating out the massive fields
indeed gives \wi.  The branch with $\epsilon _L\epsilon _R=-1$ has
$W=\half mM^{N_fN_f}$.  As in \iss , because the Kahler potential is
everywhere smooth in $M$, this branch does not give a supersymmetric
ground state.  Therefore, the branch with $\epsilon _L\epsilon _R=-1$ is
properly eliminated from the effective low energy theories with $N_f\leq
N_c- 5$.

Had there been new massless states somewhere on the branch with
$\epsilon _L\epsilon _R=-1$, the addition of the $Q^{N_f}$ mass term,
$W_{tree}$, would have led to additional ground states.  Since there
are no such extra ground states for $N_f<N_c-4$, we indeed conclude
that the manifold of quantum vacua must be smooth and without any new
massless fields.

The physics at $\ev{M}=0$ is interesting.  Classically, there were
massless quarks and gluons there.  Quantum mechanically, only the $M$
quanta are massless.  This clearly signals the confinement of the
elementary degrees of freedom.  However, as is clear from the
discussion above, the global chiral symmetry $SU(N_f) \times U(1)_R$
is clearly unbroken.  This is another example of the phenomenon
observed in \refs{\nati,\swii,\iss} of confinement without chiral
symmetry breaking.

\subsec{$N_f=N_c-3$; two branches and massless composites}

The expectation values $\ev{Q}$ generically break $SO(N_c)$ to
$SO(3)$.  The superpotential \wsymm\ can be found by examining the
limit where the first $N_f-1$ eigenvalues of $\ev{M}$ are large,
breaking the theory to $SU(2)_L\times SU(2)_R$ with one quark
$Q^{N_f}$.  Matching the running gauge coupling at the scales of the
Higgs mechanism, the scales of the
low energy $SU(2)_L\times SU(2)_R$ theory are
$\Lambda _{L,1}^5=\Lambda _{R,1}^5=\Lambda
_{N_c,N_c-3}^{2N_c-3}/ \det\widehat M$, where $\det\widehat M=\det
M/M^{N_fN_f}$.  The
expectation value $\ev{Q^{N_f}}$ breaks the $SU(2)_L\times SU(2)_R$
gauge group of this low-energy theory to a diagonally embedded $SO(3)$
with a scale $\Lambda _D^6=4\Lambda _{L,1}^5\Lambda _{R,1}^5(M^{N_fN_f})^{-2}$.
Gaugino condensation in the unbroken $SO(3)$ generates a
superpotential $W_D=2\epsilon \Lambda _D^3$, where $\epsilon =\pm 1$.
In addition, as in \ads, an instanton in the broken\foot{Typically
when the gauge group $G$ is broken to a non-Abelian subgroup $H$ along
the flat direction we do not need to consider instantons in the broken
part of the group.  The reason for that is that the phrase
``instantons in the broken part of the group'' is not well defined;
these instantons can be rotated into $H$.  Then, the strong dynamics
in the low energy $H$ gauge theory is stronger than these instanton
effects.  However, when the instantons in the broken part of the group
are well defined, their effect must be taken into account when
integrating out the massive gauge fields.  This is the case when $G$
(or one of its factors) is completely broken or broken to an Abelian
subgroup, or when the index of the embedding of $H$ in $G$ is larger
than one.  In our case the index of the embedding is 2 and therefore
we should include these instantons.} $SU(2)_L$ generates a
superpotential $W_L=2\Lambda _{L,1}^5/M^{N_fN_f}$ and an instanton in the
broken $SU(2)_R$ generates a superpotential $W_R=2\Lambda_{R,1}^5/
M^{N_fN_f}$.  Adding these three contributions and using the above
matching relations, the superpotential \wsymm\ for $SO(N_c)$ with
$N_f=N_c-3$ quarks is
\eqn\wiii{W=4(1+\epsilon ) {\Lambda ^{2N_c-3}_{N_c,N_c-3}\over \det
M}.}
The low energy theory again has two physically inequivalent phase
branches labeled by the sign of $\epsilon$.  The branch with $\epsilon
=1$ is the continuation of \wi\ to $N_f=N_c-3$.  The branch with
$\epsilon =-1$ has vanishing superpotential\foot{This happens as a result of
cancellation between a high energy contribution (the term proportional
to $1$ in \wiii) and a low energy contribution (the term proportional to
$\epsilon=-1$ in \wiii).  Can such a cancellation between high energy
and low energy contributions, which does not follow from any symmetry,
be relevant to the problem of the cosmological constant?} and,
therefore, has a quantum moduli space of vacua.

Upon adding a mass term $W_{tree}=\half mM^{N_fN_f}$ and integrating out
$Q^{N_f}$, the branch of \wiii\ with $\epsilon =1$ properly gives the
two ground states of the $\epsilon _L\epsilon _R=1$ branch of \wii.
Upon adding $W_{tree}=\half mM^{N_fN_f}$ to the $\epsilon =-1$ branch of the
theory, we must likewise get the two ground states of the $\epsilon
_L\epsilon _R=-1$ branch of \wii.  In order for the $\epsilon =-1$ branch
of the theory to not be eliminated upon adding $W_{tree}$, there must
be additional massless fields at the origin.  Since they should not
be present at generic points on the moduli space, there must be a
superpotential giving them a mass away from $M=0$.  The simplest way
to achieve that is to have fields, $q_i$, coupled to $M$ with a
superpotential which behaves as
\eqn\wfmassa{W\approx  {1\over 2\mu}M^{ij}q_iq_j}
for $M\approx 0$, where $\mu$ is a dimensionful scale needed if $q$
has dimension one because $M$ has dimension two.  Adding $W_{tree}$ to
\wfmassa\ and integrating out the massive fields indeed gives two
physically equivalent ground states with $W=0$, associated with the two
sign choices in $\ev{q_{N_f}}=\pm i\sqrt{m\mu }$.  These two ground
states correspond to the two choices of $\epsilon _1$ and $\epsilon
_2$ in the $\epsilon _1\epsilon _2=-1$ branch of the low energy
$N_f=N_c-4$ theory.

In order for \wfmassa\ to respect the global flavor symmetry, the field
$q_i$ should have the $SU(N_f)\times U(1)_R$ quantum numbers $(\overline
N_f)_{1+{1\over N_f}}$.  The most general invariant superpotential
is then
\eqn\wfmass{W={1\over 2\mu}f\left( t={(\det M) (M^{ij}q_iq_j)\over
\Lambda_{N_c,N_c-3}^{2N_c-2}}\right) M^{ij}q_iq_j.}
In order for the superpotential \wfmass\ to yield the ground states
discussed above, the function $f(t)$ must be holomorphic in a
neighborhood of $t=0$.  The $q_i$ can be rescaled to set $f(0)=1$.

The field $q_i$ was motivated by requiring the correct behavior upon
giving a flavor a mass and integrating it out.  It is a highly
non-trivial independent check that the 't Hooft anomalies with the
massless spectrum at the origin consisting of the $M^{ij}$ and $q_i$
match the anomalies \micro\ of the classical spectrum.  The fermion
component of the field $q_i$ gives the 't Hooft anomalies
\eqn\fianom{\eqalign{U(1)_R\qquad &1\cr
U(1)_R^3\qquad &N_f^{-2}\cr SU(N_f)^3\qquad &-d_3(N_f)\cr
SU(N_f)^2 U(1)_R\qquad &N_f^{-1}d_2(N_f).}}
Adding these to the contribution \manom\ of the field $M$, the
anomalies associated with the massless spectrum $M$ and $q$ do indeed
match the microscopic anomalies \micro\ for $N_f=N_c-3$.
The massless field $q_i$ is naturally identified as $q_i=\Lambda
_{N_c,N_c-3} ^{2-N_c}b_i$ where $b_i$ is the ``exotic''
composite $b_i=(Q)^{N_c-4}W_\alpha W^\alpha$, (for $N_c=4$, which we will
discuss below, this is a glueball) with the color indices contracted
with an epsilon tensor, as they have the same quantum numbers.
In terms of $b$, whose dimension is $N_c-1$, \wfmass\ is
\eqn\wfmassa{W={1\over 2 \Lambda^{2N_c-3}}f\left( t={(\det M)
(M^{ij}b_ib_j)\over \Lambda_{N_c,N_c-3}^{4N_c-6}}\right) M^{ij}b_ib_j}
where we absorbed $\mu$ in the definition of $f$.  Note that $W$ is
holomorphic in $1/\Lambda^{2N_c-3}$, the inverse of the instanton
factor.

Intuitively, one thinks of such exotics as being large and heavy bound
states.  Here we see that they become massless at $\ev{M}=0$.  This
phenomenon is similar to the massless composite mesons and baryons
found in $SU(N_c)$ theories with $N_f=N_c+1$ \nati.  Also, as with
the $N_f=N_c-4$ theories
discussed in sect. 3.2, we again see confinement without
chiral symmetry breaking.

\subsec{$N_f=N_c-2$; the Coulomb phase}

Since $M^{ij}$ is neutral under the anomaly free $U(1)_R$ symmetry, no
superpotential can be generated; the theory has a quantum moduli space
of physically inequivalent vacua labeled by the expectation values
$\ev{M^{ij}}$.  In this space of vacua the $SO(N_c)$ gauge group is
broken to $SO(2)\cong U(1)$.  Hence the theory has a Coulomb phase with
a massless photon supermultiplet.  Classically, there is a singularity
at $\det M=0$ associated with the larger unbroken gauge symmetry there.
In the quantum theory, we find a different sort of singularity at $\det
M=0$, associated with massless monopoles rather than massless vector
bosons.

The Coulomb phase of these theories can be explored by determining the
effective gauge coupling $\tau ={\theta _{eff}\over \pi}+i{8\pi \over
g_{eff}^2}$ of the massless photon on the moduli space of degenerate
vacua.  By the $SU(N_f)$ flavor symmetry, $\tau$ depends on the vacuum
$\ev{M^{ij}}$ only via the $SU(N_f)$ flavor singlet $U\equiv \det
M^{ij}$.  As in \refs{\swi -\intse}, $\tau (U, \Lambda
_{N_c,N_c-2})$ is naturally expressed in terms of a curve which can be
exactly determined by holomorphy, the symmetries, and the requirement
that $\tau$ reproduces known behavior in various understood limits.

Consider the region of the moduli space where $N_c-4$ eigenvalues of
$M^{ij}$ are large.  There the $SO(N_c)$ theory is broken to a
low-energy $SO(4)\cong SU(2)_L\times SU(2)_R$ theory with $N_f=2$.
Matching the running gauge coupling at the Higgs scales, the low energy
theory has dynamical scales given by
$\Lambda _{L,2}^4=\Lambda _{R,2}^4
=\Lambda_{N_c,N_c-2}^{2N_c-4}/U_H$,
where $U_H$ is the product of the $N_c-4$ large eigenvalues of $M^{ij}$.
The flavor singlet combination of the light
matter in the low energy $SU(2)_L\times SU(2)_R$ theory is $\hat
U=U/U_H$.  In this limit, the curve should reproduce the one
found in \intse\ for $SU(2)_L\times SU(2)_R$ with $N_f=2$:
\eqn\ivtau{y^2=x^3+x^2(-\hat U+4\Lambda _L^4+4\Lambda _R^4)+
16\Lambda _L^4\Lambda _R^4x}
(this is the curve of \intse\ upon normalizing
$M$ and $\Lambda _{L,R}$ as in sect. 2.2).   In terms of
the original high-energy theory, \ivtau\ gives (upon rescaling $x$ and
$y$)
\eqn\curve{y^2=x^3+x^2(-U+8\Lambda _{N_c, N_c-2}^{2N_c-4})+
16\Lambda _{N_c, N_c-2}^{4N_c-8}x.}

The exact curve must reproduce \curve\ in the limit where $U$ is large
compared to $\Lambda _{N_c, N_c-2}
^{2N_c-4}$.  Assuming as in \refs{\swi,\swii}
that the quantum corrections to \curve\ are polynomials in the
instanton factor $\Lambda _{N_c, N_c-2}
^{2N_c-4}$, holomorphy and the symmetries
prohibit any corrections to \curve.  Hence, the curve \curve\ is
exact.

The effective gauge coupling $\tau$ obtained from \curve\ is singular
at $U=0$ and at $U=U_1\equiv 16\Lambda ^{2N_c-4}_{N_c, N_c-2}$.  It is
found from \curve\ that, up to an overall conjugation by $T^2$, there
is a monodromy ${\cal M}_0=S^{-1}TS$ in taking $U\rightarrow e^{2\pi
i}U$ around $U=0$ and a monodromy ${\cal M}_1=(ST^{-2})^{-1}TST^{-2}$
in taking $U$ around $U_1$.  This singular behavior reveals the
presence of massless magnetic monopoles (or dyons) for vacua
$\ev{M^{ij}}=M^*$ with $\det M^*=0$ or $\det M^*=U_1$.  Note that the
spaces of such singular vacua $M^*$ are non-compact.

The number of massless monopoles (or dyons) in a singular vacuum $M^*$
follows from the monodromy of $\tau$ upon taking $M$ around $M^*$.  We
first consider vacua $M^*$ with $U=U_1$.  Taking $M$ around such an
$M^*$, $(M-M^*)\rightarrow e^{2\pi i}(M-M^*)$, takes $U-U_1\rightarrow
e^{2\pi i}(U-U_1)$ and thus gives the monodromy ${\cal M}_1$.  This
monodromy is associated with a single pair of monopoles (or dyons)
$E^\pm$, of magnetic charge $\pm 1$, with a superpotential
\eqn\wmonim{W=(U-U_1)\left[ 1+ \CO  \left( {U-U_1 \over
\Lambda_{N_c, N_c-2}^{2(N_c-2)}} \right) \right] E^+E^-.}
Away from $U=U_1$ the monopoles are massive.  At $U=U_1$ they become
massless and the photon gauge coupling is, therefore, singular.

The effective superpotential \wmonim\ properly describes the theory
only in the vicinity of the moduli space of vacua with $\det M=U_1$,
where the monopoles $E^\pm$ are light.  In particular, it is not valid
in the vicinity of $\det M=0$, where another set of monopoles are
light.  The spectrum of light monopoles near $U=0$ is more interesting
than the single light monopole of \wmonim\ near $U=U_1$.  As above,
the light spectrum of monopoles follows from considering the monodromy
implied by the curve \curve\ around a singular vacuum.  Consider
taking $M$ around a vacuum $M^*$ with $\det M^*=0$, $(M-M^*)
\rightarrow e^{2\pi i}(M-M^*)$.  This takes $U\rightarrow e^{2\pi
i(N_f-r)}U$, where $r$ is the rank of $M^*$, and thus gives the
monodromy ${\cal M}_0^{N_f-r}$.  Therefore, there must be $N_f-r$
pairs of massless monopoles in a vacuum $\ev {M}=M^*$ with $M^*$
of rank $r$.  This behavior corresponds to having $N_f$ pairs of
monopoles $q_i^+$ and $q_i^-$, of magnetic charge $\pm 1$, with a
superpotential
\eqn\wmoniim{W={1\over 2\mu}f\left( t={\det M\over
\Lambda^{2(N_c-2)}}\right) M^{ij}q_i^+q_j^-,}
with $f(t)$ holomorphic around $t=0$ and normalized so that $f(0)=1$,
to give rank $(M)$ of the monopoles a mass.  As in \wfmass, the scale
$\mu $ was introduced because $M$ has dimension two and $q$ has
dimension one.  In order for the superpotential \wmoniim\ to respect
the global flavor symmetry, the monopoles $q_i^\pm$ must have the
$SU(N_f)\times U(1)_R$ quantum numbers $(\overline N_f)_1$.

It follows from \curve\ that the massless magnetic particles at the
singularities $q_i^\pm$ and $E^\pm$ have electric charges and global
quantum numbers compatible with the identification $q^\pm_i Q^i \sim
E^\pm$.  The monopoles and dyons are visible in the semiclassical
regime of large $|U|=|\det M|$.  For $M$ proportional to the identity
matrix the global symmetry is broken to $SO(N_f)\times U(1)_R$.  We
expect to find states which are $SO(N_f)$ singlets and states which
are $SO(N_f)$ vectors (and perhaps others).  The electric charges of
these states are determined up to an even integer associated with the
monodromy $U \rightarrow e^{2\pi i}U$ ($\theta \rightarrow \theta +
2\pi$).  By shifting $\theta $ by $\pi$ the electric charges are
shifted by one unit.  A more detailed analysis of the semiclassical
spectrum can determine all the quantum numbers of the states subject
to some conventions; we did not perform such an analysis. In what
follows we will refer to the massless particles at the origin,
$q_i^\pm$, as magnetic monopoles and to the massless particles at
$\det M = 16\Lambda^{2N_c-4}_{N_c,N_c-2}$, $E^\pm$, as dyons.

The $N=1$ photon
field strength  $\CW_\alpha$  can be given
a gauge invariant description
on the moduli space in terms of the fundamental fields as
\eqn\walg{\CW_\alpha \sim W_\alpha (Q)^{N_c-2}}
where both the color and the flavor indices are contracted
antisymmetrically.  This relation will be generalized in the
non-Abelian Coulomb phase discussed in the next section.

At the origin $\ev{M}=0$, the monopoles $q_i^\pm$ are all massless.  The
massless spectrum at the origin also consists of the fields $M^{ij}$ and
the photon supermultiplet.  This spectrum, obtained from the
monodromies of \curve,
satisfies two non-trivial consistency checks.  First, at
$\ev M =0$ the full $SU(N_f)\times U(1)_R$ global symmetry is unbroken
and the 't Hooft anomalies of this massless spectrum must match the
anomalies \micro\ of the classical spectrum.  The fermion components of
the $q_i^\pm$ and the photino give the combined anomalies
\eqn\eeanom{\eqalign{U(1)_R\qquad &1\cr
U(1)_R^3\qquad &1\cr
SU(N_f)^3\qquad &-2d_3(N_f)\cr
SU(N_f)^2U(1)_R\qquad &0.}}
Adding these to the contributions \manom\ of the field $M$, the
anomalies do indeed match the microscopic anomalies \micro\ for
$N_f=N_c-2$.

Another check is to verify that, upon adding the term $W_{tree}=\half
mM^{N_fN_f}$ and integrating out the field $Q^{N_f}$, we properly
reproduce our description of $N_f=N_c-3$ discussed in the previous
subsection.  The equations of motion in the low energy effective
theory with $W_{tree}$ added lock the theory to be on a branch with
$\det M=U_1$ or on a branch with $\det M=0$.  On the branch with $\det
M=U_1$, the equations of motion obtained upon adding $W_{tree}$ to
\wmonim\ give $\ev{E^+E^-}=-m/2\det\widehat M$, where $\widehat M$ are
the mesons for the remaining $N_c-3$ light flavors.  The non-zero
expectation values of $\ev{E^\pm}$ lift the photon and confine
electric charges.  In fact, since $E^\pm$ are dyons, this phenomenon
is oblique confinement \refs{\thooft,\cardyrabin}.  The remaining
superpotential is
\eqn\wuir{W=\half mM^{N_fN_f}=8{m\Lambda _{N_c, N_c-2}^{2N_c-4}\over \det
\widehat M}.}
Using the matching relation between the high energy scale $\Lambda
_{N_c,N_c-2}$ and the scale $\Lambda _{N_c, N_c-3}$ of the low energy
theory, \wuir\ corresponds to the $\epsilon =1$ branch of \wiii.

Another branch is found by adding $W_{tree}=\half m M^{N_fN_f}$ to
\wmoniim.  The classical equations of motion show that
$\ev{q_{N_f}^\pm}\not= 0$ and hence the magnetic $U(1)$ is Higgsed.
This is confinement of the original electric variables.  The
non-trivial function $f(t)$ in
\wmoniim\ and the constraint from the $U(1)$ D-term make the
explicit integration out of the massive modes complicated.  However,
it is easy to see that only $\widehat M^{\hat i \hat j}$ with $\hat i,
\hat j=1,...,N_f-1$ and $q_{\hat i}= {1 \over 2 \sqrt{m\mu}} (q_{\hat
i}^+q_{N_f}^- -q_{\hat i}^-q_{N_f}^+)$ remain massless.  (This
expression for $q_{\hat i}$ is the gauge invariant interpolating field
for the massless component of $q_{\hat i}^\pm$.)  Their effective
superpotential is
\eqn\effsupno{W= {1\over 2\mu}\hat f \left( \hat t = {(\det \widehat M)
( \widehat M^{\hat i \hat j} q_{\hat i} q_{\hat j}) \over m
\Lambda _{N_c, N_c-2}^{2(N_c-2)}}\right)
\widehat M^{\hat i \hat j} q_{\hat i} q_{\hat j}}
where $\hat f(\hat t)$ depends on $f(t)$ in \wmoniim.  (The condition
{}from the $U(1)$ D-term is important in showing that a non-trivial $f(t)$
leads to a non-trivial $\hat f(\hat t)$).   This branch thus yields
the $\epsilon =-1$ branch of the low energy $N_f=N_c-3$ theory, as
described by \wfmass.

In the $N=2$ theory of \swi\ there are also massless monopoles and
dyons which lead to confinement when they condense.  In that case the
confining branch and the oblique confinement branch are related by a
global $Z_2$ symmetry.  Therefore, there is no physical difference
between them.  In some of the examples in \swii\ there are massless
monopoles and dyons which are not related by any global symmetry.
When they condense they lead to confinement and oblique confinement.
However, since these theories have matter fields in the fundamental
representation of the gauge group, there is no invariant distinction
between Higgs, confinement and oblique confinement \higgscon\
in these examples.  In the present cases the Higgs, confining and
oblique confinement branches are physically inequivalent.

We see here an interesting physical phenomenon. Upon giving $Q^{N_f}$
a mass, some of the magnetic monopoles $q_i^\pm$ condense, leading to
confinement, and the remaining massless monopoles are interpreted as
massless exotics (or glueballs).  A similar phenomenon was observed in
$SU(N_c)$ theories in \sem\ where massless magnetic quarks became
massless baryons.  We conclude that this phenomenon is generic; some
of the gauge invariant composites (baryons, glueballs, exotics) can be
thought of as ``magnetic.''

\newsec{$N_c\geq 4$, $N_f\geq N_c-1$; magnetic $SO(N_f-N_c+4)$ gauge
theory}

As discussed in \sem, the infra-red behavior of these theories has a
dual, magnetic description in terms of an $SO(N_f-N_c+4)$ gauge theory
with $N_f$ flavors of dual quarks $q_i$
and the additional gauge singlet field
$M^{ij}$.  For $N_c-2< N_f \le {3 \over 2} (N_c-2)$ the
magnetic degrees of freedom are free in the infra-red while for ${3
\over 2} (N_c-2)< N_f < 3 (N_c-2)$ the electric and the magnetic
theories flow to the same non-trivial fixed point of the renormalization
group. Although the two theories are different away from the extreme
infra-red, they are completely equivalent at long distance.  This means
that the two (super) conformal field theories at long distance are
identical, having the same correlation functions of all of the operators,
including high dimension (irrelevant) operators.

The fields in the magnetic theory have the anomaly free $SU(N_f)\times
U(1)_R$ charges
\eqn\duasot{\eqalign{
q &\qquad  (\overline N_f)_{{N_c-2 \over N_f}} \cr
M &\qquad (\half N_f(N_f+1))_{2{(N_f-N_c+2)\over N_f}}  \cr}}
and a superpotential
\eqn\typsup{W={1\over 2\mu}M^{ij}q_i\cdot q_j}
(an additional term is required for $N_f=N_c-1$).  The scale $\mu$ is
needed for the following reason.  In the electric description
$M^{ij}=Q^i \cdot Q^j$ has dimension two at the UV fixed point and
acquires some anomalous dimension at the IR fixed point.  In the
magnetic description $M$ is an elementary field of dimension one at the
UV fixed point.  Denote it by $M_m$.  In order to relate it to $M$ of
the electric description a scale $\mu$ must be introduced with the
relation $M=\mu M_m$.  Below we will write all the expressions in
terms of $M$ and $\mu$ rather than in terms of $M_m$.

For generic $N_c$ and $N_f$ the scale of the magnetic theory, $\tilde
\Lambda$, is related to that of the electric theory, $\Lambda$, by
\eqn\mgscrgen{\Lambda^{3(N_c-2)-N_f}\tilde
\Lambda^{3(N_f-N_c+2)-N_f}=C(-1)^{N_f-N_c}\mu ^{N_f}}
where $C$ is a dimensionless constant which we will determine below
and $\mu$ is the dimensionful scale explained above.  This relation of
the scales has several consequences:

\noindent
1. It is easy to check that it is preserved under mass deformations and
along the flat directions (more details will be given below).  The phase
$(-1)^{N_f-N_c}$ is important in order to ensure that this is the case.

\noindent
2. It shows that as the electric theory becomes stronger the magnetic
theory becomes weaker and vice versa.

\noindent
3. Because of the phase $(-1)^{N_f-N_c}$, the relation \mgscrgen\ does
not look dual -- if we perform another duality transformation it becomes
$\Lambda^{3(N_c-2)-N_f}\tilde \Lambda^{3(N_f-N_c+2)-N_f}=C(-1)^{N_c}
\tilde \mu ^{N_f}$ and therefore
\eqn\mumutil{\tilde \mu=-\mu .}
This minus sign is important when we dualize again.  The dual of the
dual magnetic theory is
an $SO(N_c)$ theory with scale $\Lambda$, quarks $d^i$, and
additional singlets $M^{ij}$ and $N_{ij}=q_i \cdot q_j$, with superpotential
\eqn\typsupa{W={1 \over 2\tilde \mu} N_{ij}d^i \cdot d^j+{1 \over 2\mu}
M^{ij}N_{ij}={1 \over 2 \mu} N_{ij}(M^{ij}-d^i \cdot d^j) .}
The first term is our standard superpotential of duality transformations
(as pointed out in \sem\ the relative minus sign between it and
\typsup, which follows from \mumutil, is common in Fourier or Legendre
transforms).  The second term is simply copied from \typsup.  $M$ and
$N$ are massive and can be integrated out using their equations of
motion $N=0$, $M^{ij}=d^i\cdot d^j$.  This last relation shows that
the quarks $d$ can be identified with the original electric quarks
$Q$.  The dual of the magnetic theory is the original electric theory.

\noindent
4.  Differentiating the action with respect to $\log \Lambda$
relates the field strengths of the electric and the magnetic theories
as $W_\alpha^2 = -\tilde W_\alpha^2$.  The minus sign in this expression
is common in electric magnetic duality, which maps $E^2-B^2=-(\tilde
E^2-\tilde B^2)$.  In our case it shows that the gluino bilinear in the
electric and the magnetic theories are related by
$\lambda\lambda=-\tilde \lambda\tilde \lambda$.

The electric theories are also invariant under a discrete $Z_{2N_f}$
symmetry generated by $Q \rightarrow e^{2\pi i \over 2N_f} Q$ and
charge conjugation $\CC$.  In the dual description these are generated
by $q \rightarrow e^{-2\pi i \over 2N_f} \CC q$ and $\CC$
respectively.  Note that the $Z_{2N_f} $ symmetry commutes with the
electric gauge group but does not commute with the magnetic one.  This
is similar to the action of the parity operator $P$ in Maxwell theory:
in the dual description $P$ is replaced with $P\CC$.

The gauge invariant (primary) chiral operators of the electric theory
are
\eqn\sonop{\eqalign{
M^{ ij }&=\half Q^iQ^j \cr
B^{[i_1,...,i_{N_c}]}&=Q^{i_1}...Q^{i_{N_c}} \cr
b^{[i_1,...,i_{N_c-4}]}&=W_\alpha ^2 Q^{i_1}...Q^{i_{N_c-4}} \cr
\CW_\alpha^{[i_1,...,i_{N_c-2}]}&=W_\alpha Q^{i_1}...Q^{i_{N_c-2}} \cr
}}
with the gauge indices implicit and contracted.  These operators get
mapped to gauge invariant operators of the magnetic theory as
\eqn\emopmap{\eqalign{M^{ij}&\rightarrow M^{ij}\cr
B^{[i_1\dots i_{N_c}]}&\rightarrow \epsilon ^{i_{1}\dots i_{N_f}}\tilde
b _{i_{N_c+1}\dots i_{N_f}}\cr
b^{[i_1\dots i_{N_c-4}]}&\rightarrow
\epsilon ^{i_1\dots i_{N_f}}\tilde B_{[i_{N_c-3}\dots i_{N_f}]}\cr
\CW_\alpha^{[i_1,...,i_{N_c-2}]}&\rightarrow \epsilon ^{i_1\dots
i_{N_f}}(\tilde{\CW}_\alpha)_{[i_{N_c-1},...,i_{N_f}]},\cr}}
where $\tilde B$, $\tilde b$, and $\tilde\CW_\alpha$ are the magnetic
analogs of the operators in \sonop.  The last of these relations has
already been noted in \walg.  Note that these maps are consistent with
both the continuous and the discrete symmetries.

\subsec{$N_f=N_c-1$; magnetic $SO(3)$ gauge theory}

The dual magnetic description is in terms of an $SO(3)$ gauge theory
with $SO(3)$ quarks $q_i$ in $(\overline N_f)_{{N_c-2\over N_f}}$ of
the global $SU(N_c)\times U(1)_R$, $SO(3)$ singlets $M^{ij}$ with
$SU(N_c)\times U(1)_R$ quantum numbers as before, and a superpotential
\eqn\wmncmi{W={1\over 2\mu}M^{ij}q_i\cdot q_i-{1\over 64
\Lambda _{N_c, N_c-1}^{2N_c-5}}\det M,}
where $\mu$ is the dimensionful normalization factor discussed in the
introduction to this section.  The scale $\tilde \Lambda _{3, N_c-1}$
of the magnetic $SO(3)$ theory with $N_f=N_c-1$ massless quarks is
related to the scale of the electric theory by
\eqn\mgiiisc{2^{14}(\Lambda _{N_c, N_c-1}^{2N_c-5})^2\tilde
\Lambda _{3, N_c-1}^{6-2(N_c-1)}=\mu ^{2(N_c-1)}.}
Since this relation is like the square of \mgscrgen, the phase discussed
there is not present.  The normalizations of the second term in \wmncmi\
and of the relation \mgiiisc\ are determined for consistency of the
various deformations of the theory (see below).  Since $N_f\geq 3$, the
magnetic $SO(3)$ gauge theory is not asymptotically free and is,
therefore, free in the infra-red.  At the infra-red fixed point the free
fields $q_i$ and $M$ all have dimension one.  Hence, the superpotential
\wmncmi\ is irrelevant and the infra-red theory has a large accidental
symmetry.  However the superpotential \wmncmi , including the $\det M$
term, is essential in order to properly describe the theory when
perturbed by mass terms or along flat directions.  Also, without the
$\det M$ term, the magnetic theory \wmncmi\ would have a $Z_{4N_f}$
symmetry in contrast to the $Z_{2N_f}$ symmetry \disc\ of the electric
theory.  Using the symmetries and holomorphy around $M=q=0$, it is easy
to see that, unlike \wfmass\ or \wmoniim, \wmncmi\ cannot be modified by
a non-trivial function of the invariants.

At the origin $\ev{M}=0$, the fields $M^{ij}$, $q_i$ and the $SO(3)$
vector bosons are all massless.  The $SU(N_c)\times U(1)_R$ anomalies
associated with this massless spectrum match the anomalies \micro\ of
the classical spectrum \sem.

\bigskip
\centerline{\it Flat directions}

We now consider the moduli space of vacua in the dual magnetic
description, verifying that it agrees with the moduli space of vacua
in the original electric description.  For $M\not=0$, \wmncmi\ gives
the magnetic quarks a mass matrix $\mu^{-1}M$.  The low energy theory
is the magnetic $SO(3)$ with $k=N_f- \rank (M)$ massless dual quarks
$q$.  For $\rank (M)=N_f$ all the dual quarks are massive.  Then, using
\massmatch\ in the magnetic theory, the low energy, pure gauge,
$SO(3)$ Yang-Mills theory has a scale $\tilde \Lambda _{3,0}^6 =\tilde
\Lambda _{3,N_f}^{6-2(N_c-1)}\det (\mu^{-1}M)^2$.  Gluino condensation
in the magnetic $SO(3)$ leads to two vacua, $\ev{\tilde
W_\alpha^2}=\epsilon \tilde \Lambda _{3,0}^3$ with $\epsilon =
\pm 1$, and hence an additive term $2\ev{\tilde W_\alpha^2}=2\epsilon
\tilde \Lambda _{3,N_f}^{6-2(N_c-1)}\mu ^{-N_c+1}\det M$
in the superpotential.  Adding this to the term proportional to $\det
M$ in \wmncmi\ and using \mgiiisc, the $\epsilon =1$ branch reproduces
the moduli space of
supersymmetric ground states with generic $\ev{M}$.  The only massless
fields on this moduli space of generic $\ev{M}$ are the components of
$M$.  The $\epsilon =-1$ branch will be interpreted in sect. 6.

For $\rank (M) = N_f-1 $ the low energy theory is the magnetic $SO(3)$
with one massless flavor, which we take to be $q_{N_f}$.  This is a
magnetic version of the theory analyzed in \swi.  It has a massless
photon and massless monopoles at $\ev{u}\equiv \ev{q_{N_f}^2}
=4 \epsilon\tilde \Lambda ^2_{3,1}$ ($\epsilon=\pm1$) where, using
\mgiiisc\ and \massmatch , the scale $\tilde \Lambda
_{3,1}$ of the low energy theory is given by $\tilde \Lambda
^4_{3,1}=2^{-14}(\mu \Lambda ^{5-2N_c}\det\widehat M)^2$, with
$\det\widehat M=\det M/M^{N_fN_f}$ the product of the $N_f-1$ non-zero
eigenvalues of $M$.  The low energy superpotential near the massless
monopole points $u\approx 4\epsilon \tilde \Lambda _{3,1}^2$ is
\eqn\wmagmag{W={1\over 2\mu}M^{N_fN_f}(u-{1\over 32\Lambda
_{N_c,N_c-1}^{2N_c-5}}\mu \det\hat M)-{1\over
2\mu}(u-4\epsilon\tilde \Lambda_{3,1}^2)\tilde E^+_{(\epsilon )}\tilde
E^-_{(\epsilon )}}
where the normalization of the $E^+_{(\epsilon )}\tilde E^-_{(\epsilon )}$
term was arranged for convenience.  The $M^{N_fN_f}$ equations of
motion give $\ev{u}=2^{-5}\Lambda _{N_c,N_c-1}^{5-2N_c}\mu \det\hat M$,
fixing the
magnetic theory to the $\epsilon =+1$
supersymmetric ground state with a massless
monopole $\tilde E^\pm_{(+)}$ in addition to the massless photon.  The $u$
equation of motion gives $M^{N_fN_f}=\tilde E^+_{(+)}\tilde E^-_{(+)}$.  The
monopole $\tilde E^\pm_{(+)}$ is magnetic relative to the magnetic $SO(3)$
variables; it is electric in terms of the original electric variables.
Indeed, using the electric theory we easily see that a flat direction
with $\rank (M)=N_f-1=N_c-2$ breaks the $SO(N_c)$ gauge group to
$SO(2) \cong U(1)$ with one of the elementary quarks which is charged
under this $U(1)$ remaining massless.  In the magnetic description we
find it as a massless collective excitation.  This interpretation is
strengthened by the relation $M^{N_fN_f}=\tilde E^+_{(+)}\tilde E^-_{(+)}$.

For $\rank (M) \le N_f-2$ the low energy theory is $SO(3)$ with
$k=N_f-\rank (M) \ge 2$ flavors which is either free or is at a
non-trivial fixed point (only for $k=2$) of the beta function.  It is
dual to the answer one gets in the electric variables.

\bigskip
\centerline{\it Mass deformations}

Adding a $Q^{N_f}$ mass term $W_{tree}=\half mM^{N_fN_f}$ to the
magnetic theory \wmncmi , the $M^{N_fN_f}$ equation of motion gives
$q_{N_f}^2=2^{-5}\mu\Lambda _{N_c, N_c-1}^{5-2N_c}\det
\widehat M-\mu m$; this generically breaks the magnetic $SO(3)$ gauge
group to $SO(2)$.  Integrating out the massive fields, the low energy
magnetic $SO(2)$ theory has neutral fields $ M^{\hat i\hat j}$ and
fields $q_{\hat i}^\pm $ of $SO(2)$ charge $\pm 1$, where $\hat
i=1\dots N_c-2$, along with a superpotential $W_{tree}={1\over 2\mu}
M^{\hat i\hat j}q^+_{\hat i}q^-_{\hat j}$.  Instantons in the broken
magnetic $SO(3)$ can generate additional terms\foot{As in the
discussion of footnote 2, these terms should be included because, when
the magnetic $SO(3)$ is broken to $SO(2)$, there are well-defined
instantons in the broken part of the gauge group.}, modifying the
superpotential to $W={1\over 2\mu} f(\det\widehat M/\Lambda _{N_c,
N_c-2}^{2(N_c-2)})M^{\hat i\hat j}q^+_{\hat i}q^-_{\hat j}$.  This is
the theory \wmoniim , with the $q _i^\pm$ becoming the monopoles of
the theory with $N_f=N_c-2$.  Therefore, the $SO(3)$ gauge group of
\wmncmi\ really deserves to be called ``magnetic.''

There is another special point on the moduli space of vacua.  For
large $\det\widehat M$ the first $N_f-1$ of the $q_i$ should be
integrated out and the low energy theory is the magnetic $SO(3)$
theory with the quark $q_{N_f}$ and scale $\tilde \Lambda _{3,1}
^4=2^{-14}(\mu \Lambda _{N_c, N_c-1}^{5-2N_c}\det\widehat M)^2$.
The term $\half M^{N_fN_f}u$ in the superpotential, where
$u=q_{N_f}^2$, locks the magnetic theory to be at one of the vacua
where the theory has massless monopoles.  Near these two vacua the low
energy theory is described by
\eqn\ubyupm{W \approx {1\over 2\mu}M^{N_fN_f}(u-{1\over 32 \Lambda
_{N_c,N_c-1}^{2N_c-5}}\mu \det\widehat M+
\mu m)-{1\over 2\mu}(u-4\epsilon \tilde \Lambda _{3,1}^2)
\tilde E^+_{(\epsilon)}\tilde E^-_{(\epsilon)},}
where the fields $\tilde E ^\pm_{(\epsilon )}$, which appear as the
result of strong coupling phenomena in the magnetic theory, can be
interpreted as a monopole or dyon of that theory.  Using the equations
of motion, there is no supersymmetric vacuum for $\epsilon =1$ and the
low energy theory for the $\epsilon =-1$ vacuum is
\eqn\smonl{W\approx {1 \over 2} (m-{1\over 16\Lambda _{N_c, N_c-1}
^{2(N_c-2)-1}}\det\widehat M)\tilde E^+_{(-)}\tilde E ^-_{(-)}.}
This theory corresponds to \wmonim, with $\tilde E^\pm_{(-)}$ the dyons
which are massless at $\det\widehat M=16\Lambda _{N_c, N_c-2}^{2N_c-4}$.

Adding more masses, we gradually reduce $N_f$.  The monopoles or
dyons condense and lead to confinement or oblique confinement.  For
fewer than $N_c-4$ massless flavors the confining branch disappears
and there is only the oblique confinement branch with
\eqn\obbra{W_{oblique}= -{1 \over 32\Lambda_{N_c,N_c-1}^{2N_c-5}}
\det M,}
which is the continuation of \wi\ to $N_f=N_c-1$.  The superpotential
\obbra\ does not mean that the flat directions of the massless theory
are lifted.  As in \intse, it should be only used to reproduce
$\ev{M}$ when the quarks are massive.  It is present only in the
oblique confinement branch and not in the Higgs branch.

To conclude, the monopoles $q_i^\pm$ at the origin of the $N_f=N_c-2$
theories have a weakly coupled magnetic description in terms of the
components $q_i^\pm$ of the quarks in the dual theory.  The massless
dyons $ E^\pm$ at $\det\widehat M =16\Lambda _{N_c, N_c-2}^{2N_c-4}$
appear strongly coupled both in the original electric description and
in terms of the dual magnetic description.

\subsec{$N_f=N_c$; magnetic $SO(4)$ gauge theory}

The electric $SO(N_c)$ theory with $N_f=N_c$ quarks has the gauge
invariant ``baryon'' operator $B=\det Q$ in addition to the ``mesons''
$M^{ij}=Q^i\cdot Q^j$.  As discussed in sect. 2, the classical moduli
space of vacua is constrained by $B=\pm \sqrt{\det M}$.

As discussed in \sem, these theories have a dual magnetic description
in terms of an $SO(4)\cong SU(2)_L\times SU(2)_R$ gauge theory with
$N_f$ flavors of quarks $q_i$ in the $(2,2)$ dimensional
representation of $SU(2)_L\times SU(2)_R$ along with the $SO(4)$
singlets $M^{ij}$ and the superpotential
\eqn\wnfncmqq{W={1\over 2\mu} M^{ij}q_i\cdot q_j.}
As in the previous subsection, the symmetries and holomorphy around
$M=q=0$ uniquely determine this superpotential; unlike \wfmass\ or
\wmoniim, \wnfncmqq\ cannot be modified by a non-trivial function.

The scales $\tilde \Lambda _{s, N_c}$ of the magnetic $SU(2)_s$
are equal and are related to the electric scale by
\eqn\mfscrln{2^8\tilde\Lambda _{s, N_c}^{6-N_c}\Lambda _{N_c,
N_c}^{2N_c-6}=\mu ^{N_c} \qquad\hbox{for $s=L,R$}.}
The magnetic $SU(2)_L\times SU(2)_R$ theory is not asymptotically free
for $N_f>5$.  Therefore, for $N_f>5$ the magnetic theory is free in the
infra-red.  For $N_c=N_f=4,5$ the theory is asymptotically free and has
an interacting fixed point with the quarks $q_i$ and the $SU(2)_L\times
SU(2)_R$ gauge fields in an interacting non-Abelian Coulomb phase.  It
is dual to the electric description in terms of the original $SO(N_c)$
theory with $N_f=N_c$ quarks $Q^i$ in a non-Abelian Coulomb phase.

The theory has an anomaly free $SU(N_f)\times U(1)_R$ global symmetry
with the fields $q_i$ in the $(\overline N_f)_{{N_c-2\over N_c}}$ and the
$M^{ij}$ in the $(\half N_f(N_f+1))_{{4\over N_c}}$.  At $\ev{M}=0$ the
fields $q_i$ and $M^{ij}$ are all massless.  Since the global
$SU(N_f)\times U(1)_R$ symmetry is unbroken at the origin, the 't Hooft
anomalies of this massless spectrum must match the classical anomalies
\micro; they do indeed match \sem.

\bigskip
\centerline{\it Flat directions}

Consider the theory in a vacuum of non-zero $\ev{M}$.  The $M$
equations of motion give $q_i\cdot q_j=0$; the $SO(4)$ D--terms imply
that the only solution is $\ev{q_i}=0$.  The low energy theory around
this point is the magnetic $SO(4)$ with $k=N_f- \rank (M)$ dual quarks
$q$.  For $\rank (M)= N_f$ there are no light dual quarks and the
low-energy magnetic theory is $SU(2)_L\times SU(2)_R$ Yang-Mills
theory with, using \massmatch\ and \mfscrln ,
the scales $\tilde \Lambda _{s,0}^6=2^{-8}\det M \Lambda _{N_c,
N_c}^{6-2N_c}$ for $s=L,R$.  There is gaugino condensation in the
$SU(2)_s$: $\ev{(\tilde W_\alpha )^2_s}=\epsilon _s\tilde \Lambda
_{s,0}^3$ for $s=L,R$; $\epsilon _s=\pm 1$ label four vacua.  This
leads to a superpotential $W=2(\epsilon _L+\epsilon _R)\tilde \Lambda
^3$.  The two vacua with $\epsilon _L\epsilon _R=1$ have $W\approx \pm
{1\over 4} \Lambda ^{3-N_f}(\det M)^{1/2}$ and do not lead to
supersymmetric vacua.  The two vacua with $\epsilon _L\epsilon _R=-1$
give a branch with $W=0$; each gives a supersymmetric ground state.
We thus find that there are two vacua for $\rank (M) = N_f$,
corresponding to the sign of $\ev{(\tilde W_\alpha )^2_L-(\tilde
W_\alpha )^2_R}$.  This is in agreement with the classical moduli
space of vacua discussed in sect. 2 with, by the identification \sem\
$B\sim (\tilde W_\alpha )^2_L-(\tilde W_\alpha )^2_R$, the two vacua
for $\ev{M}$ of rank $N_c$ corresponding to the sign of
$B=\pm\sqrt{\det M}$.

For $\rank (M)=N_f-1$, the low energy theory is the magnetic $SO(4)$
theory with one flavor, $q_{N_f}$.  As discussed in sect. 3.3, it has no
massless gauge fields and a massless composite $\tilde q$ appears.  In
this case, the massless composite is a glueball $(\tilde W_\alpha
)^2_L-(\tilde W_\alpha )^2_R$.  The effective Lagrangian is $W={1\over
2\mu}N_{N_fN_f}(M^{N_fN_f}- \tilde q^2)$; integrating out
$N^{N_fN_f}=q_{N_f}\cdot q_{N_f}$ gives $M^{N_fN_f}=\tilde q^2$.  The
metric is smooth in terms of $\tilde q$ rather than $M^{N_fN_f}$.  The
field $\tilde q$ of the magnetic theory can be seen semiclassically in
the electric theory.  For $\rank (M)=N_f-1=N_c-1$ the electric theory is
completely Higgsed but one of the quarks remains massless.  Its gauge
invariant interpolating field is $B=\det Q$, which is indeed mapped under
the duality to the massless glueball $(\tilde W_\alpha )^2_L-(\tilde
W_\alpha )^2_R$ of the magnetic theory.

For $\rank (M)=N_f-2$ the low energy theory is the magnetic $SO(4)$ with
two flavors discussed in sect. 3.4,
which is in the Coulomb phase with massless
magnetic monopoles.  These can be seen in the electric theory as being
some of the components of the elementary quarks, which are charged under
the unbroken electric $U(1)$ for $\rank (M)=N_f-2=N_c-2$.

For $\rank (M)=N_f-3$ the low energy theory is the magnetic $SO(4)$ with
three flavors discussed in sect. 4.1.  It is in a free non-Abelian
magnetic phase with gauge group $SO(3)$ with three flavors of magnetic
quarks.  These are precisely the electric degrees of freedom of the
underlying $SO(N_c)$ theory, which is Higgsed along the flat directions
with $\rank (M)=N_f-3$ to an electric $SO(3)$ subgroup.  Here we see
these elementary quarks and gluons appearing out of strong coupling
dynamics in the dual magnetic theory.

For $\rank (M) \le N_f-4$ the low energy theory is the magnetic
$SO(4)$ with more than three flavors.  It is either at a non-trivial
fixed point of the beta function or not asymptotically free.

We see that for $\rank (M) < N_f$ there is a unique ground state which
can be interpreted either in the electric or in the magnetic theory.

\bigskip
\centerline{\it Mass deformations}

Now consider perturbing the theory by $W_{tree}=\half mM^{N_cN_c}$ to
give a mass to the $N_c$-th electric quark.  Adding $W_{tree}$ to
\wnfncmqq, the $M^{N_cN_c}$ equation of motion gives $\ev{q_{N_c}^2}=-
\mu m$, which breaks the magnetic $SU(2)_L\times SU(2)_R$ to the
diagonal $SU(2)_D$.  The $q_{N_c}$ equations of motion give $M^{
iN_c}=0$ and the $M^{\hat iN_c}$ equations of motion give $q_{\hat
i}\cdot q_{N_c}=0$ for $\hat i=1\dots N_c-1$.  The remaining low
energy theory is the diagonal magnetic $SO(3)$ gauge theory with
$N_c-1$ triplets $\hat q_{\hat i}$ and the $SO(3)$ singlets $M^{\hat
i\hat j}$, where $\hat i$, $\hat j=1\dots N_c-1$.  These fields have a
superpotential coming from \wnfncmqq.  In addition, there is a
contribution to the superpotential associated with instantons in the
broken magnetic $SU(2)_L$ and $SU(2)_R$.  In particular, for $\det
\widehat M\neq 0$, the superpotential \wnfncmqq\ gives masses to the
first $N_c-1$ dual quarks $q_{\hat i}$.  The low energy theory has one
dual quark $q_{N_c}$ and the $SU(2)_s$ scales are $\tilde \Lambda
_{s,1}^5= 2^{-8}\mu \det\widehat M \Lambda _{N_c, N_c}
^{6-2N_c}$.  Instantons in the broken magnetic $SU(2)_s$ generate the
superpotential $W_{inst}=2(\tilde \Lambda _{L,1}^5+\tilde \Lambda
_{R,1}^5)/q_{N_c}\cdot q_{N_c}$.  Using $\ev{q_{N_c}\cdot q_{N_c}}
=-m\mu$ and $m\Lambda _{N_c, N_c}^{2N_c-6}=\Lambda _{N_c,
N_c-1}^{2N_c-5}$,
\eqn\winstin{W_{inst}=-{1\over 64} \Lambda _{N_c,
N_c-1}^{5-2N_c} \det\widehat M.}
Combining $W_{inst}$ with the
superpotential coming from \wnfncmqq , the low energy magnetic theory
properly yields the magnetic $SO(3)$ theory with $N_c-1$ flavors and
the superpotential \wmncmi\ discussed in the previous subsection.  Using
\massmatch\ and \higgsmatch\ in the electric and magnetic theories, the
scale relation \mfscrln\ properly yields the scale relation \mgiiisc\
for the low-energy electric and magnetic theories.

\subsec{$N_f>N_c$; magnetic $SO(N_c-N_c+4)$ gauge theory}

The superpotential in the dual magnetic description is
\eqn\wsoem{W={1\over 2\mu}M^{ij}q_i\cdot q_j.}
As in the previous subsections, symmetries and holomorphy around
$M=q=0$ uniquely determine this superpotential; a non-trivial
function as in \wfmass\ or \wmoniim\ cannot be present.  The scale
of the magnetic theory is related to that of the electric theory by
\eqn\mgscrln{2^8\Lambda _{N_c,N_f}^{3(N_c-2)-N_f}\tilde \Lambda
_{N_f-N_c+4,N_f}^{3(N_f-N_c+2)-N_f}=(-1)^{N_f-N_c}\mu ^{N_f}.}

The one-loop beta function of $SO(N_f-N_c+4)$ with $N_f$ quarks
reveals that the magnetic theory is not asymptotically free for
$N_f\leq {3\over 2}(N_c-2)$.  For this range of $N_f$, the magnetic
gauge theory is free in the infra-red and provides a weakly coupled
description of the strongly coupled electric theory.  For ${3\over
2}(N_c-2)<N_f<3(N_c-2)$, the magnetic theory is asymptotically free
and has an interacting fixed point with the $SO(N_f-N_c+4)$ theory in
a non-Abelian Coulomb phase.  For this range of $N_f$, there is also
an electric description in terms of the original $SO(N_c)$ theory with
$N_f$ quarks in a non-Abelian Coulomb phase.  The magnetic
description is at stronger coupling as $N_f$ is increased and the
electric description is at weaker coupling.  For $N_f\geq 3(N_c-2)$,
the magnetic description is at infinite coupling whereas the electric
description is free in the infra-red.

At the origin of $\ev{M}$ the fields $M^{ij}$, the $q_i$, and the
$SO(N_f-N_c+4)$ vector bosons are all massless. The 't Hooft anomalies
of this massless spectrum match the anomalies \micro\ of the classical
theory \sem.

\bigskip
\centerline{\it Flat directions}

Now consider the theory along the flat directions of non-zero $\ev{M}$.
$k=N_f- \rank (M)$ of the $q_i$ remain massless.  The $F$ and $D$ terms
of the dual theory fix $\ev{q_i}=0$.  For $\rank (M) >N_c$, there is no
supersymmetric ground state at $\ev{q_i}=0$ because a superpotential
analogous to \wi\ is generated in the magnetic theory.  For $\rank (M)
=N_c$, the low energy magnetic theory is analogous to the theory
considered in sect. 3.2; there are two supersymmetric ground states at
the origin corresponding to the two sign choices for $\epsilon _L$ in
the $\epsilon _L\epsilon _R=-1$ branch of the magnetic analog of \wii.
The same is also true in the underlying electric theory.  For $\rank
(M)=N_c-1$ the low energy theory is $SO(N_f-N_c+4)$ with $k=N_f-N_c+1$.
It is analogous to the theory considered in sect. 3.3.  It has no
massless gauge fields but massless composites.  These can be interpreted
as some of the components of the elementary electric quarks.  For $\rank
(M) =N_c-2$ the low energy magnetic theory is $SO(N_f-N_c+4)$ with
$N_f-N_c+2$ massless flavors.  It is analogous to the theory discussed
in sect. 3.4.  This magnetic theory has a massless photon which is at
infinite coupling because of the massless magnetic monopoles at the
origin, $\ev{q_i}=0$.  This gives the dual description of the fact which
is obvious in the electric variables: that there is a massless photon
with massless charged elementary quarks when $\rank (M)= N_c-2$.  For
${3 \over 2} N_c - {1 \over 2} N_f -3 \le \rank (M) < N_c-2$ the low
energy theory is still strongly coupled.  It is dualized as in this
section to a free electric theory $SO(N_c- \rank (M))$ with $N_f-\rank
(M)$ massless quarks.  Again, this result is obvious in the original
electric variables.  For $\rank (M) < {3 \over 2} N_c - {1 \over 2} N_f
-3 $ the magnetic degrees of freedom are either interacting or free.

To summarize, the moduli space of supersymmetric vacua is given by
the space of $\ev{M}$ of rank at most $N_c$ along with an additional
sign when $M$ is of rank $N_c$.  We thus recover the classical moduli
space, discussed in sect. 2, of the electric theory in terms of strong
coupling effects in the magnetic description.  Conversely, some of the
strong coupling phenomena of the previous subsections can be
understood from the classical moduli space of the dual magnetic
theory.

\bigskip
\centerline{\it Mass deformations}

Adding a $Q^{N_f}$ mass term, $W_{tree}=\half mM^{N_fN_f}$, to the
electric theory gives a low energy electric $SO(N_c)$ theory with
$N_f-1$ massless quarks.  Adding $W_{tree}$ to the theory \wsoem , the
$M^{N_fN_f}$ equations of motion give $\ev{q_{N_f}}\neq 0$, breaking
the magnetic $SO(N_f-N_c+4)$ gauge theory with $N_f$ quarks to
$SO(N_f-N_c+3)$ with $N_f-1$ quarks.  The $q_{N_f}$ equations of
motion give $M^{\hat iN_f}=0$ and the $M^{\hat iN_f}$ equations of
motion give $q_{\hat i} \cdot q_{N_f}=0$ for $\hat i=1\dots N_f-1$.
The remaining low energy theory is then a magnetic $SO(N_f-N_c+3)$
theory with $N_f-1$ flavors and the superpotential \wsoem.  This low
energy magnetic theory is, indeed, the magnetic dual to the low energy
$SO(N_c)$ theory with $N_f-1$ massless quarks.  Using \massmatch\ and
\higgsmatch\ in the electric and magnetic theories, the relation
\mgscrln\ in the high-energy theory implies that the scales of the
low-energy theory are also related as in \mgscrln.
When we flow from an electric theory with $N_f=N_c+1$
to the electric theory with $N_f=N_c$, the magnetic $SO(5)$ with
$N_f=N_c+1$ is broken to the magnetic $SU(2)_L\times SU(2)_R$ of the
previous section with $N_f=N_c$.

Another way to analyze the theory with mass terms is to consider the
massless theory for generic values of $M$.  The dual quarks acquire mass
${1 \over \mu} M$ and the low energy magnetic theory is a pure gauge
$SO(N_f-N_c+4)$ with scale $\tilde \Lambda_L^{3(N_f-N_c+2)}=
\mu^{-N_f}\tilde\Lambda_{N_f-N_c+4,N_f}^{3(N_f-N_c+2)-N_f} \det
M$.  Gluino condensation in this theory leads to an effective
superpotential
\eqn\effsupgen{\eqalign{
W_{eff}&=\half (N_f-N_c+2)2^{4/(N_f-N_c+2)}\tilde \Lambda_L^3\cr
&=\half (N_f-N_c+2) \left(16
\mu^{-N_f}\tilde\Lambda_{N_f-N_c+4,N_f}^{3(N_f-N_c+2)-N_f} \det
M\right)^{1/(N_f-N_c+2)}\cr
&=\half (N_f-N_c+2) \left((-1)^{N_f-N_c} {16
\Lambda_{N_c,N_f}^{3N_c-6-N_f} \over \det M }\right)^{-1/
(N_f-N_c+2)}\cr
&= \half (N_c-N_f-2)\left( {16 \Lambda_{N_c,N_f}^{3N_c-6-N_f}\over
\det M}\right)^{1/(N_c-N_f-2)},}}
which is the same as the continuation of \wi\ to these values of $N_c$,
$N_f$.  This guarantees that the expectation values of $\ev{M^{ij}}$ are
reproduced correctly when mass terms are added to the magnetic theory.

\newsec{$SO(3)$}

In this section we discuss the case $N_c=3$, which exhibits some new
phenomena.  As before, the dual of $SO(3)$ with $N_f$ quarks, $Q^i$,
is an $SO(N_f+1)$ theory with $N_f$ dual quarks, $q_i$.  The $N_c=3$
theory is invariant under the discrete $Z_{4N_f}$ symmetry \disc.  As
in the previous section, the $Z_{2N_f}$ subgroup acts in the magnetic
theory as $q\rightarrow e^{-2\pi i/2N_f}\CC q$; the full $Z_{4N_f}$
should be generated by the ``square root'' of this operation.  The
correct ``square root'' of the charge conjugation $\CC$ is, as we will
show, the $SL(2,Z)$ electric-magnetic duality modular transformation
$A=TST^2S$.  Therefore, the $Z_{4N_f}$ symmetry is realized
non-locally in the dual theories. In other words, the discrete
$Z_{4N_f}$ symmetry is a ``quantum symmetry'' in the dual description.

For $N_c=3$ a new term has to be added to the dual theory, proportional to
\eqn\newterm{\det (q_i \cdot q_j).}
This term can be motivated by several arguments.  One of them is by
considering the dual of the dual.  As we discussed in sect. 4.1, the
dual of $SO(N_f+1)$ with $N_f$ flavors is $SO(3)$ with $N_f$ flavors
with an extra interaction term proportional to $\det M$.  The term
\newterm\ is then needed to ensure that the dual of the dual \wmncmi\
is the original theory.  For $N_f \ge 3$ this determines its
coefficient:
\eqn\newtermc{W={1\over 2\mu}M^{ij}q_i\cdot q_j+
{1 \over 64 \tilde \Lambda_{N_f+1,N_f}^{2(N_f-1)-1}}\det
(q_i \cdot q_j),}
where
\eqn\latillamt{\epsilon 2^7 \tilde \Lambda_{N_f+1,N_f}^{2(N_f-1)-1}
\Lambda_{3,N_f}^{3-N_f}=(-1)^{3-N_f}\mu^{N_f}.}
$\epsilon=\pm 1$ arises from taking the square root of the instanton
factor $\Lambda_{3,N_f}^{3-N_f}$ of the electric $SO(3)$ theory and
the phase $(-1)^{3-N_f}$ keeps the relation \latillamt\ preserved
along the flat directions and with mass perturbations.  The superpotential
\newterm\ is not renormalizable; this will be discussed below.

The term \newterm\ is invariant under all the continuous symmetries of
the electric theory.  However, both magnetic $SO(N_f+1)$ instantons
(except for $N_f=1,2$) and \newterm\ break some of the discrete
symmetries.  Only charge conjugation $\CC$ and the $Z_{2N_f}$ subgroup
of the $Z_{4N_f}$ symmetry remain unbroken.  The underlying $Z_{4N_f}$
symmetry seems to be explicitly broken.  The naive symmetry
transformation flips the sign of \newterm\ and shifts the theta angle of
the magnetic theory (for $N_f\not= 1,2$) by $\pi$.  This is consistent
with the coefficient in \newtermc\ and the relation \latillamt\ as
written in terms of the instanton factor $\tilde
\Lambda_{N_f+1,N_f}^{2(N_f-1)-1}$ of the magnetic theory.
We would like to interpret this as
follows.  The original electric $SO(3)$ theory has, in fact, two dual
descriptions corresponding to the two signs of this term and a shift
of theta by $\pi$ (for $N_f\not= 1,2$).  One of them is ``magnetic,''
which was discussed as the ``electric'' theory in sect. 4.1.  The
other dual theory is ``dyonic.''  It will be discussed further in
section 6.  These two theories are related by another duality
transformation, which extends the group of $N=1$ duality
transformations to $SL(2,Z)$.  More precisely, we have only $S_3 \cong
SL(2,Z)/\Gamma(2)$, which permutes these three theories.  The full
$Z_{4N_f}$ symmetry includes the modular transformation which
exchanges the magnetic and dyonic theories -- it appears as a quantum
symmetry in the dual description.

We will now discuss these theories in more detail starting with small
values of $N_f$.

\subsec{$N_f=1$; Abelian Coulomb phase and quantum symmetries}

This is the $N=2$ theory discussed in \swi.  Since no superpotential
is compatible with the anomaly free $U(1)_R$ symmetry, the theory has
a quantum moduli space of vacua labeled by the expectation value of
the massless meson field\foot{Our convention for the normalization of
$\Lambda ^2_{3,1}$ differs by a factor of 2 from that of \swi\ and the order
parameter $u$ of \swi\ satisfies $u=\half M$.}  $M= Q^2$.
The $SO(3)$ gauge symmetry is broken to $SO(2)\cong U(1)$ on this
moduli space so the theory has a Coulomb phase with a massless photon,
similar to the generic case of $N_f=N_c-2$.  The effective gauge
coupling of the photon is given by the curve
\eqn\swicurve{y^2=x^2(x-M)+4\Lambda _{3,1}^4x.}

As discussed in \swi, there is a massless magnetic monopole $q_{(+)}^\pm$
at $M=4\Lambda _{3,1}^2$ and a massless dyon $q_{(-)}^\pm$ at
$M=-4\Lambda _{3,1}^2$.
Their effective superpotentials are
\eqn\effsoto{W_\pm =f_\pm (M/\Lambda^2_{3,1})q_{(\pm )}^+ q_{(\pm )} ^-}
where in the notation of \swi, $f_+=a_D(M/\Lambda _{3,1}^2)$ and
$f_-=ia_D(M/\Lambda _{3,1}^2)+ ia(M/\Lambda^2_{3,1})$, satisfying
\eqn\ztonf{f_+(-M/\Lambda^2_{3,1})=f_-(M/\Lambda^2_{3,1}).}
$f_\pm$ is holomorphic around $M=\pm 4\Lambda^2_{3,1}$ and has a cut along
$[\mp 4\Lambda _{3,1}^2, \infty)$.  Expanding around $M \approx \pm
4\Lambda^2_{3,1}$,
the superpotentials are
\eqn\fpmex{W_\pm \approx {1\over 2\mu}\left(M  \mp 4\Lambda _{3,1}^2\right)
q_{(\pm )}^+q_{(\pm )}^-}
where the first term is our standard $Mq^+q^-$ term and the second term
is \newterm.

This theory provides an example of a quantum symmetry.  The theory has
the global symmetry group $((SU(2)_R\times Z_8^R)/Z_2)\times \CC$,
where the $Z_8^R$ is an $R$ symmetry whose generator, $R$, acts on all
of the $N=2$ super charges as a $e^{2\pi i/8}$ phase and $\CC$ is
charge conjugation.  (The $Z_{4N_f}=Z_4$ discussed in the introduction
of this section is embedded in $SU(2)_R\times Z_8^R$.)  The $Z_8^R$
generator acts on the scalar component of $M$ as $R: M\rightarrow -M$
and is therefore broken for $M\neq 0$.  Since $R^2$ acts as charge
conjugation on the squarks, the $Z_8^R$ symmetry is spontaneously
broken to a $Z_4^R$ generated by $R^2\CC$ for $M\neq 0$
(alternatively, we could combine the generator of this symmetry with
the broken Weyl transformation in the gauge group).  At $M=0$ the full
$Z_8^R$ symmetry is restored.  What is not obvious is that its
generator includes an $SL(2,Z)$ modular transformation, $w=RA$ with
$A=(TS)^{-1}S(TS)$ \swi.  Since $A^2=\CC$, $w^2=R^2\CC$ generates the
$Z_4$ found away from $M=0$.  The necessity of the modular
transformation $A$ in $w$ can be seen, for example, by considering the
central term in the $N=2$ algebra, $Z=an_e+a_Dn_m$ \swi.  Since the
generator $R$ in $w$ multiplies the $N=2$ charge by $e^{2\pi i/8}$,
$Z$ must transform under $w$ as $w:Z\rightarrow iZ$.  It is easily
seen from the integral expressions for $a(M)$ and $a_D(M)$ \swi\ that
$an_e'+a_Dn_m'=i(an_e+a_Dn_m)$ at $M=0$ if $n_e'$ and $n_m'$ are
related to $n_e$ and $n_m$ by the modular transformation
$A=(TS)^{-1}S(TS)$.  Note that $A=\CC T(S^{-1}T^2S)$; thus, $A$ is
congruent to $\CC T$ modulo multiplication by the monodromy
$S^{-1}T^2S$ associated with looping around one of the singularities.

For $M\neq 0$ the broken generator $w$ maps, for example, the massless
monopole at $M=4\Lambda _{3,1}^2$ to the massless dyon at
$M=-4\Lambda^2_{3,1}$.
At the origin, where the $Z_8^R$ symmetry is restored, these states are
degenerate and are mapped to one another by the symmetry.  Since the
fields which create these two particles are not relatively local, it
is impossible for $w$ to be given a local realization.  Indeed, it is
a modular transformation.  Furthermore, since $A$ cannot be
diagonalized by an $SL(2,Z)$ transformation, there is no photon field
which is invariant under it.  Therefore, $A$ cannot be realized
locally even in the low energy effective Lagrangian at $M=0$ which
includes only the photon multiplet.

We can now interpret the superpotentials \effsoto\ as reflecting the
symmetries along the lines of the general comments in the introduction
to this section and the discussion of the term \newterm.  The electric
$SO(3)$ theory has two dual theories.  One of them, which we can refer
to as the ``magnetic dual,'' describes the physics around
$M=4\Lambda^2_{3,1}$ with the superpotential $W_+$ in \effsoto.  The other
dual, which can be called the ``dyonic dual'' is valid around
$M=-4\Lambda^2_{3,1}$
and is described by $W_-$ in \effsoto.  The magnetic
dual is related to the underlying electric theory by the
transformation $S$ in $SL(2,Z)$ (modulo $\Gamma (2)$) while the dyonic
dual is related to the electric description by the $SL(2,Z)$
transformation $ST$ (again, modulo $\Gamma (2)$).

Below we will see more complicated examples of quantum symmetries and of
magnetic and dyonic duals of the same electric theory.

\subsec{$N_f=2$; non-Abelian Coulomb and quantum symmetries}

As discussed in \intse, this theory has three branches described by the
superpotentials
\eqn\wehc{W=e{\det M\over 8\Lambda _{3,2}}+\half \Tr mM,}
where $e=0, \pm 1$ label the three branches.  The branch with $e=0$ is
appropriate for the Higgs or Coulomb phases of the theory.  These
phases are obtained for $\det m=0$.  For $m=0$ the generic point in
the moduli space is in the Higgs phase. When only $m_{22}$ is non-zero
the low energy theory is that discussed in \swi.  It has a massless
monopole point at $M^{11}=4m_{22}\Lambda_{3,2}$ and a massless dyon
point at $M^{11}=-4m_{22}\Lambda_{3,2}$ (there is an arbitrary choice
here in which one is magnetic and which is dyonic).  When $\det m\neq
0$, the monopole (or the dyon) condenses and leads to confinement (or
oblique confinement).  Corresponding to these phenomena there are two
branches of the theory: a confining branch with $e=-1$ in \wehc\ and
an oblique confinement branch with $e=1$ in \wehc.

The theory has two dual descriptions in terms of an $SO(3)$ theory with
$N_f=2$:
\eqn\sottda{W= {2\over 3\mu}\Tr Mq\cdot q
+\epsilon \left( {8\tilde \Lambda _{3,2} \over 3 \mu^2}\det M
+{1\over 24\tilde \Lambda _{3,2}}\det q\cdot q \right),}
where $\epsilon=\pm 1$ labels the two duals and the scales of the
theories are related by
\eqn\mtiiscrln{64\Lambda _{3,2}\tilde \Lambda _{3,2}=\mu^2.}
A $\det M$ term, as in \sottda , was present in the previously discussed
$N_f=N_c-1$ cases and the $\det q \cdot q$ term is as in \newterm.  In
\mtiiscrln\ we took the square root of a relation involving the
instanton factors, $\Lambda^2 _{3,2}$ and $\tilde \Lambda^2 _{3,2}$, of
the two groups.  The sign ambiguity in doing so is represented in
\sottda\ by $\epsilon$.  The coefficients in \sottda\ and \mtiiscrln\
are fixed to guarantee the duality.  We will later also determine them
by flowing down from other theories.

The $\det q\cdot q$ term in \sottda\ is not renormalizable.
This term can be replaced with ${2\over
3\mu}\Tr Lq\cdot q-{32\tilde \Lambda _{3,2}\epsilon \over 3\mu
^2}\det L$, which yields the $\det q\cdot q$
term in \sottda\ upon integrating out $L$.  The theory with the field
$L$ included has the superpotential
\eqn\renw{W={2\over 3\mu}\Tr (M+L)q\cdot q +{8\epsilon
\tilde\Lambda _{3,2}\over 3\mu ^2}(\det M-4\det L),}
which is renormalizable.

The electric theory has a $Z_8$ symmetry, generated by $Q\rightarrow
e^{2\pi i/8}Q$, and charge conjugation $\CC$.  In the magnetic theory,
the $Z_8$ symmetry of the electric theory takes $M\rightarrow e^{2\pi
i/4}M$ and $q\rightarrow e^{-2\pi i/8}A q$, where $A$ is a non-local
transformation such that $A^2=\CC$.  We do not have an explicit
expression for $A$ but the consistency of our answers suggests that it
exists and hence the $Z_8$ is a quantum symmetry.

Just as in the electric theory \wehc, the magnetic theory also has three
branches with the superpotentials
\eqn\sottd{W= {2\over 3\mu}\Tr MN
+\epsilon \left( {8\tilde \Lambda _{3,2} \over 3 \mu^2}\det M
+{1\over 24\tilde \Lambda _{3,2}}\det N \right)
+\tilde e{\det N\over 8\tilde \Lambda _{3,2}},}
where $N_{ij}\equiv q_i\cdot q_j$ and, as above, $\tilde e=0,
\pm 1$ labels the three branches.

The superpotential \sottd\ is quadratic in both $M^{ij}$ and
$N_{ij}$.  Therefore, $M^{ij}$ or $N_{ij}$ can be
integrated out.  Integrating out $N_{ij}$ yields
\eqn\wsottnon{W_{\tilde e}={8 \tilde \Lambda _{3,2} \over \mu^2 }
\left ( {\tilde e - \epsilon \over 1+3\tilde e \epsilon} \right)\det M+
\half \Tr mM
={1 \over 8  \Lambda _{3,2} }
\left ( {\tilde e - \epsilon \over 1+3\tilde e \epsilon} \right)\det M+
\half \Tr mM,}
where we added the $Q^i$ mass terms $W_{tree}=\half \Tr mM$.  This is
the same as \wehc\ with
\eqn\eetild{e= {\tilde e - \epsilon \over 1+3\tilde e \epsilon}.}
$\tilde e=0$ describes the weakly coupled Higgs branch of the dual
theories.  It leads to $e = -\epsilon$, which corresponds to the two
strongly coupled branches of the electric theory.  The Higgs branch of
the $\epsilon=1$ theory describes the confining branch of the electric
theory ($e=-1$) while the Higgs branch
of the $\epsilon=-1$ theory describes the oblique confinement branch
of the electric theory ($e=1$).  Therefore, we can refer to the
$\epsilon=1$ theory as magnetic and to the $\epsilon=-1$ theory as
dyonic.  The two other branches of the dual theories are strongly
coupled.  The branches with $\tilde e=\epsilon$ (oblique confinement
of the magnetic theory and confinement of the dyonic theory) lead to
$e=0$ and therefore to the Higgs branch of the electric theory.
Similarly, the branches with $\tilde e=-\epsilon$ (confinement of the
magnetic theory and oblique confinement of the dyonic theory) give
another description of the strongly coupled branches of the electric
theory.

This discussion leads to a new interpretation of the first term in
\wehc.  In the electric theory this term appears as a consequence of
complicated strong coupling dynamics in the confining and the oblique
confinement branches of the theory.  In the dual descriptions it is
already present at tree level.

An equivalent analysis can be performed with the renormalizable theory
\renw.  For example,  along the flat
directions $q$ gets an effective mass ${4\over 3\mu}(M+L)$ and can be
integrated out.  The low energy theory is pure-gauge magnetic $SO(3)$
Yang-Mills theory.  The low energy effective superpotential is
\eqn\wsottlnon{W_{eff}={8\epsilon \tilde \Lambda _{3,2}
\over 3\mu ^2}(\det M-4\det L)+{32\eta\tilde \Lambda _{3,2}
\over 9\mu ^2}\det (M+L),}
where $\eta=\pm 1$.  The first terms in \wsottlnon\
are the tree-level terms of \renw\ and the last
term is generated by gaugino condensation in the magnetic $SO(3)$
Yang-Mills theory.  Integrating out $L$, we find
\eqn\wrennolgc{W={8\tilde \Lambda _{3,2}\over \mu ^2}\left({\eta\epsilon
+1\over 3\epsilon -\eta}\right)\det M.}
Therefore, the flat direction is obtained for $\eta=-\epsilon$.

To conclude, we have three equivalent theories: electric, magnetic and
dyonic.  Every one of them has three branches: Higgs, confinement and
oblique confinement.  The map between the branches of the different
theories is an $S_3$ permutation described by \eetild.

We now consider the Coulomb phase of the electric theory, obtained by
adding $W_{tree}=\half mM^{22}$ to \sottd\ and integrating out the
massive fields by their equations of motion.  This gives
\eqn\mtteqs{\eqalign{{2 \over 3 \mu} q_2 \cdot q_2 +{8\epsilon \tilde
\Lambda _{3,2} \over 3 \mu^2} M^{11}+{1\over 2}m&=0
\cr q_1 \cdot q_2&=0}\qquad
\eqalign{M^{22}&=-{\epsilon\over 16\tilde \Lambda _{3,2}}
\mu ^{-1}q_1 \cdot q_1 \cr M^{12}&=0.}}
The expectation value of $q_2$ breaks the gauge group to $SO(2)$ for
$M^{11}+3 \epsilon m \mu^2/16 \tilde \Lambda _{3,2}\neq 0$.  The
remaining charged fields, $q_1^+$ and $q_1^-$, couple through the low
energy superpotential
\eqn\sottii{{1\over 2\mu}(M^{11}-m{\epsilon \mu ^2\over 16\tilde
\Lambda _{3,2}})q_1^+q_1^- = {1\over 2\mu}(M^{11}-4\epsilon m \Lambda
_{3,2})q_1^+q_1^-.}
This superpotential is corrected by contributions from
instantons in the broken magnetic $SO(3)$ theory.  For large $m$
their contribution is small and can be ignored.  We see that the
theory has massless fields $q_1^\pm$ at $M^{11}=4\epsilon m\Lambda
_{3,2}=4\epsilon \Lambda _{3,1}^2$.
The massless fields for $\epsilon=1$ ($\epsilon=-1$) can be
interpreted as the monopoles (dyons) of the $N_f=1$ theory \swi.  We see
them as weakly coupled states in the $\epsilon=1$ ($\epsilon=-1$)
theory.  This is in accord with the interpretation of the $\epsilon=1$
($\epsilon=-1$) theory as magnetic (dyonic).

The other monopole point of the $N_f=1$ theory arises from strong
coupling dynamics in the dual theories.  To see that, note that the
analysis above is not valid for $\epsilon M^{11}+12m\Lambda_{3,2}
\approx 0$, where the mass of $q_1$ is above the Higgs expectation
value of $q_2$.  In that case, $q_1$ should be integrated out first.
For $u\equiv q_2^2\neq 0$, the effective mass of $q_1$ is ${4M^{11} \over
3\mu} +{\epsilon \over 12\tilde \Lambda _{3,2}}u$ and the
scale of the low energy magnetic theory is thus $\tilde \Lambda
_{3,1}^4= ({4\over 3\mu}\tilde \Lambda _{3,2} M^{11}+{\epsilon
\over 12 } u)^2$.  There are massless monopoles at $u=\pm 4 \tilde
\Lambda_{3,1}^2$, i.e.\ at $u =16\tilde \Lambda _{3,2}\mu ^{-1}
M^{11}/(\pm 3 - \epsilon )$.  The $M^{22}$ equation of motion in
\mtteqs\ gives $\mu ^{-1}u=-(\epsilon /16\Lambda _{3,2})M^{11}-(3/4)m$
and therefore $M^{11}=4m\Lambda _{3,2}{\mp 3 +\epsilon \over 1\pm
\epsilon}$.  For non-zero $m$ every value of $\epsilon$ leads to only
one solution, at $M^{11}=-4\epsilon m\Lambda _{3,2}=-4\epsilon \Lambda
_{3,1}^2$.  We have thus found the other monopole of the $N_f=1$ theory as a
result of strong coupling dynamics in the dual theories.

An analysis similar to the one above for leads to a strongly coupled
state in the dual theories along the flat directions with $\det M=0$
in the $m=0$ case.  This state can be interpreted as the massless
quark of the electric theory at that point.

Consider taking the dual of the dual theories \sottda.  The result is
an $SO(3)$ theory with scale $\tilde{\tilde \Lambda}_{3,2}=\Lambda
_{3,2}$, $N_f=2$ quarks $d^i$, gauge singlet fields $M$ and
$N$, and a superpotential
\eqn\sottdad{\eqalign{W&={2\over 3\mu}\Tr MN+\epsilon
\left({1\over 24\Lambda _{3,2}}\det M+{1\over 24\tilde
\Lambda_{3,2}}\det N\right)\cr
&-{2\over 3\mu}\Tr Ndd+\eta\left({1\over 24\tilde \Lambda _{3,2}}\det
N+{1\over 24\Lambda _{3,2}}\det d\cdot d\right),}}
where, $\epsilon =\pm 1$ and $\eta=\pm 1$ label the different duals.
Using \mtiiscrln , the first line in \sottdad\ is the superpotential
associated with the duals \sottda\ and the
second line are from the duals of that.  When $\epsilon=-\eta$, $N$ is a
Lagrange multiplier implementing the constraint $M=d\cdot d$ and the
superpotential is $W=0$.  These duals are identified as the original
electric theory with $d^i=Q^i$.  On the other hand,
the duals \sottdad\ with $\epsilon =\eta$, upon integrating out $N$,
have a superpotential
\eqn\sottdadm{W={\epsilon \over 12\Lambda _{3,2}}M^{ij}(d\cdot
d)_{ij}-{\epsilon \over 24\Lambda _{3,2}}\left(\det M+\det d\cdot
d\right),}
where we define $(d\cdot d)_{ij}\equiv \epsilon _{ik}\epsilon
_{jl}d^k\cdot d^l$.  These appear to be new dual theories.  However,
this is not the case.  In particular, defining $q_i\equiv \epsilon
_{ij}\sqrt{\epsilon}(\Lambda _{3,2}/\tilde \Lambda _{3,2})^{1/4} d^i$
and using \mtiiscrln , the superpotential \sottdadm\ is equivalent to
the magnetic dual superpotential \sottda.  In addition, upon scaling
{}from $d^i$ to $q_i$, the anomaly changes the scale of the theory from
$\Lambda _{3,2}$ to $\tilde \Lambda _{3,2}$.  The duals in \sottdadm\
are, therefore, equivalent to the magnetic duals  \sottda.  To
summarize, $SO(3)$ with $N_f=2$ has three descriptions: the original
electric one and the two magnetic duals of \sottda.  Taking duals of
the duals permutes these three descriptions.

\subsec{$N_f=3$}

The theory has a bare coupling constant $\tau _0={\theta_0\over
2\pi}+{4\pi\over g_0^2}i$ which is not renormalized at one loop.  With
$W_{tree}=0$ the two loop beta function makes the theory not
asymptotically free and therefore it is free in the infra-red.

There are dual magnetic and dyonic theories with gauge group
$SO(4)\cong SU(2)_L\times SU(2)_R$ with $N_f=3$ flavors $q_i$ and a
superpotential
\eqn\sofsup{W={1\over 2\mu}\Tr Mqq + {1 \over 64\tilde \Lambda
_{s,3}^3}\det qq,}
where the second term is as in \newtermc\ and the scales $\tilde
\Lambda _{s,3}$ of the magnetic $SU(2)_s$ are equal and are given by
\eqn\mqfscrln{\epsilon 2^{7}e^{i\pi \tau _0}\tilde \Lambda
_{s,3}^3=\mu^3.}
$\epsilon =\pm 1$ reflects the fact that the term $e^{i\pi \tau _0}$
in \mqfscrln\ is the square-root of the $SO(3)$ instanton factor.  The
theory is magnetic or dyonic depending on the sign of $\epsilon$.  In
\sottda\ and \mtiiscrln\ $\epsilon$ appears only in the superpotential
and not in the instanton factor of the magnetic group.  Here, on the
other hand, $\epsilon$ arises in relating the instanton factor for the
magnetic $SU(2)_s$ to the square root of the instanton factor for the
electric $SO(3)$.

The analysis of the flat directions with $W_{tree}$ is similar to that
of sect. 4.2.  The $\det qq$ term in \sofsup, which was not present in
the larger $N_c=N_f$ theories considered in sect. 4.2, does not
significantly modify the analysis.

Consider the theory perturbed by the mass term $W_{tree}=\half
mM^{33}$.  The electric theory flows to $N_f=2$ with a scale
$\Lambda_{3,2}$ given by
\eqn\thrtwf{\Lambda_{3,2}^2=m^2e^{2\pi i\tau _0}. }
The threshold factor was determined using our threshold conventions and
the results of \swii.  Adding the mass term $W_{tree}=\half mM^{33}$ to
\sofsup , the equations of motion give $\ev{q_3^2}=-\mu m$, which breaks
the magnetic $SU(2)_L\times SU(2)_R$ gauge group to a diagonally
embedded $SO(3)$.  The relation \mqfscrln\ and the matching relations
\massmatch\ and \higgsmatch\ imply that the scale $\tilde \Lambda
_{3,2}$ of the low-energy magnetic theory is related to the scale
$\Lambda _{3,2}$ of the low-energy electric theory as in \mtiiscrln.
Integrating out the massive fields at tree level we find
\eqn\wnaive{W_{tree}={1\over 2\mu}\Tr \hat M \hat q \hat q
+{\epsilon \over 32\tilde \Lambda _{3,2}}\det \hat q \hat q.}
As in \winstin, we should also include the contribution of instantons
in the broken part of the magnetic gauge group.  The presence of the
non-renormalizable $\det \hat q \hat q$ term affects the instantons
contributions.  Following the discussion which led to \renw , we
replace this term with $\Tr L \hat q \hat q-32 \epsilon
\tilde \Lambda _{3,2}\det L$, which is the same upon integrating out
$L$.  We can now repeat the analysis leading to \winstin.  The
effective mass of $ \hat q_i$ is ${1 \over \mu} M + 2L$ and therefore
the low energy superpotential is
\eqn\wfixL{W={1\over 2\mu}\Tr (\hat M+2\mu L)\hat q\hat q -
32\epsilon\tilde \Lambda _{3,2}\det L
+{2\epsilon \tilde \Lambda _{3,2} \over \mu ^2}\det (\hat M+2\mu L).}
Integrating out $L$, \wfixL\ yields
\eqn\wfixnol{W={2\over 3\mu}\Tr \hat M \hat q \hat q+\epsilon\left(
{8\tilde \Lambda _{3,2}\over 3\mu ^2}\det \hat M+{1\over 24\tilde
\Lambda _{3,2}}\det \hat q \hat q \right),}
the superpotential in \sottda.

\subsec{$N_f=3$ with $W_{tree}=\beta \det Q$; $N=4$ duality as $N=1$
duality}

We now consider perturbing the electric theory by adding the cubic
superpotential $W_{tree}=\beta \det Q$.  For $\beta= \sqrt 2$ the
theory becomes the $N=4$ $SO(3)$ Yang-Mills theory.  (We normalize the
fields such that the whole Lagrangian has a prefactor of $1 \over
g_0^2$ and therefore in $N=4$ the physical gauge coupling equals the
physical Yukawa coupling.)  Because of the anomaly, as we rescale
$\beta $ to this value, $\tau_0$ changes to
\eqn\taubetax{e^{2\pi i\tau _E}=e^{2\pi i\tau_0}({\beta \over
\sqrt {2}})^4= {1 \over 4}e^{2\pi i\tau_0}\beta ^4. }
After the rescaling the kinetic term of the three $Q$'s does not have
the proper normalization.  However, it is straightforward to see that
this feature is achieved in the infra-red.  In other words, the theory
is attracted to the $N=4$ $SO(3)$ Yang-Mills theory in the infra-red
with $\tau _E={\theta\over 2\pi}+{4\pi\over g^2}i$ given by \taubetax.
An interesting consequence of \taubetax\ is that the theory only makes sense
for $|\beta| \leq \sqrt{2}|e^{-\half i\pi\tau_0}|$.

Using \emopmap , we map the electric operator $\det Q$ to $(\tilde
W_\alpha)_L^2-(\tilde W_\alpha)_R^2$ in the magnetic theory.  Therefore,
adding $W_{tree}=\beta \det Q$ to the electric theory modifies the
magnetic $SU(2)_L\times SU(2)_R$ theory to have $\tilde \Lambda _{L,3}^3
\not= \tilde \Lambda_{R,3}^3$.  The symmetries then determine that the
superpotential \sofsup\ is modified to
\eqn\sofsupp{W={1\over 2\mu}\Tr Mq\cdot q + {2 \epsilon e^{i\pi \tau_0} \over
\mu^3 } f(\tau_E, \epsilon) \det q\cdot q,}
and the scale relation \mqfscrln\ is similarly modified to
\eqn\mqfscrt{\epsilon 2^7 e^{i\pi \tau _0}\tilde \Lambda_{s,3}^3
g_s(\tau_E, \epsilon)=\mu ^3}
where $f$ and $g_s$ are functions which we do not determine except to
note that for $\beta =0$, $\tau_E= i \infty$ and $f(i\infty,
\epsilon)=g_s(i \infty, \epsilon)=1$.

In the infra-red, the magnetic $SU(2)_L\times SU(2)_R$ theory with
$\tilde \Lambda _L\neq \tilde \Lambda _R$ also flows to
an $N=4$ theory.  This can be seen by considering the $\tilde
\Lambda _L\gg \tilde\Lambda _R$ limit, corresponding to some value of
$\tau_E$ which we denote by $\tau_*$. For $\tilde\Lambda _R=0$, the
magnetic theory is $SU(2)_L$ with six doublets coupled through
\sofsupp, which breaks the global $SU(6)$ to $SU(3)\times SU(2)_R$
under which the doublets $q_i$ are in the $(\overline 3, 2)$.  The
strong $SU(2)_L$ dynamics confines them to the fields $N_{ij}=q_i\cdot
q_j$, in the $(\overline 6,1)$, and $\phi ^i$ in the $(3,3)$ of
$SU(3)\times SU(2)_R$.  As in \nati\ these fields couple through the
superpotential $-\half\tilde\Lambda _L^{-1}N_{ij}\phi^i \cdot\phi ^j+
{1\over 8}
\tilde\Lambda _L^{-3}\det N+2\det \phi $, where we rescaled $\phi^i$ to
dimension one.
Adding this to \sofsupp\ and adding a mass term $\half \Tr mM$, we
find the superpotential
\eqn\suidynxa{{1\over 2\mu}M^{ij}N_{ij}+ {2 \epsilon e^{i\pi \tau_0}
\over \mu^3 } \left[ f(\tau_*, \epsilon) + 2^3g_L(\tau_*, \epsilon)
\right] \det N  - \half \tilde\Lambda _L^{-1} N_{ij}\phi ^i\cdot \phi
^j+2\det \phi+ \half \Tr mM .}

Now we can weakly gauge $SU(2)_R$.  At energies higher than $\tilde
\Lambda_L$, its coupling constant runs with the scale $\tilde \Lambda_R$.
Below $\tilde \Lambda_L$, $SU(2)_R$ couples to the three triplets
$\phi^i$ and its Wilsonian coupling constant $\tau_R$ does not run.  It
thus satisfies
\eqn\taur{e^{2 \pi i \tau_R} \sim {\tilde \Lambda_R^3 \over \tilde
\Lambda_L^3}}
and hence for $\tilde \Lambda_R \ll \tilde \Lambda_L$, $\tau_R \approx
i\infty$.

The fields $M$ and $N$ in \suidynxa\
are massive and can be integrated out.  The
$M$ equation of motion sets $N=-\mu m$ and the superpotential
\suidynxa\ becomes
\eqn\suidynxaa{2\det \phi + \half {\mu \over \tilde\Lambda _L}
m_{ij}\phi^i\cdot \phi ^j- 2 \epsilon e^{i\pi \tau_0} \left[ f(\tau_*,
\epsilon)  +  2^3 g_L(\tau_*, \epsilon) \right]\det m.}
For $m=0$ this theory is attracted in the infra-red
to an $SU(2)$ $N=4$ theory with weak coupling $\tau_R$.

Away from the limit $\tilde \Lambda_R \ll \tilde \Lambda_L$, the
magnetic theory also flows to an $N=4$ theory with some coupling
$\tau_R$ which depends only on $\tau_E$.

It is clear that this $N=4$ theory with $\tau_R$ is not the same as
the original $N=4$ theory with $\tau_E$ given by \taubetax.  The
original one, with coupling $\tau_E$, is weakly coupled for $\beta \ll
1$, with $\tau_E \sim {2\over \pi i} \log \beta$.  The other one,
with coupling $\tau_R$, is strongly coupled for $\beta \ll 1$, where
$\tilde \Lambda_L \approx \tilde \Lambda_R$.  Conversely, the theory
with coupling $\tau_R$ is weakly coupled when $\tilde \Lambda_L \gg
\tilde \Lambda_R$, which happens for $\beta \sim 1$, where the original
theory is strongly coupled.  Although we could not prove that
$\tau_E=-1/\tau_R$, we suspect that this fact is true and we interpret
this theory as being the $N=4$ dual of the original theory.  This shows
that the $N=1$ duality of \sem\ is a generalization of the $N=4$ duality
of \om.

The meson operator $M^{ij}$
of the electric theory can be related to
the corresponding operator in the $\tau_R
\approx i \infty$ limit of the magnetic theory by
differentiating \suidynxaa\ with respect to $m$:
\eqn\massrel{M^{ij}= {\mu \over \tilde\Lambda _L} \phi^i\cdot \phi ^j
- 4 \epsilon e^{i\pi \tau_0} \left[ f(\tau_*)
+  2^3 g_L(\tau_*) \right]\det m \left({1 \over m} \right)^{ij},}
which is not simply proportional to the bilinear $ \phi^i\cdot \phi ^j$.
A similar shift was observed in the special case of $m$ with
one vanishing eigenvalue and the two other eigenvalues equal in the flow
{}from $N=4$ to $N=2$ in \swii , thus strengthening our interpretation of
the duality map.

\newsec{More Dyonic Duals}

In sect. 4.1 we found that the theory with $N_f=N_c-1$ has a dual
magnetic description in terms of an $SO(3)$ theory with $N_f$ quarks
and the superpotential \wmncmi.  We now consider the dual of this
magnetic theory.  In sect. 5 we found that $SO(3)$ theories have both
magnetic and dyonic duals.  Therefore, there are two duals of the dual
of the theory with $N_f=N_c-1$.  Both are in terms of an $SO(N_c)$
gauge theory with $N_f$ matter fields $d^i$ and gauge singlet fields
$M^{ij}$ and $N_{ij}$ with a superpotential
\eqn\wddmnd{W={1\over 2\mu }\Tr N(M-dd)-{1\over 64\Lambda
_{N_c,N_c-1}^{2(N_c-2)-1}}(\det M-\epsilon \det dd),}
with $\epsilon =\pm 1$, and scales
\eqn\ddscri{\tilde{\tilde \Lambda} _{N_c,N_c-1}^{2N_c-5}=\epsilon \Lambda
_{N_c,N_c-1}^{2N_c-5}.}
The $N$ equation of motion of \wddmnd\ gives
$M^{ij}=d^i\cdot d^j$. The theory \wddmnd\ with $\epsilon =1$, the
magnetic dual of the magnetic dual, gives $W=0$; this is the original
electric theory with $d^i$ identified with $Q^i$.  On the other hand,
the theory \wddmnd\ with $\epsilon =-1$ has a superpotential
\eqn\wdyonici{W=-{1\over 32\Lambda _{N_c,N_c-1}^{2N_c-5}}\det d\cdot
d={1\over 32 \tilde{\tilde
\Lambda}_{N_c,N_c-1}^{2N_c-5}}\det d\cdot d}
and, from \ddscri, a theta angle differing from that of the original
electric theory by a shift by $\pi$.  We will refer to this theory as
the ``dyonic dual'' of the original theory.

Near the origin in field space the operator \wdyonici\ is irrelevant
and does not affect the dynamics.  The flat directions of this theory
are more subtle.  Analyzing the theory classically we might conclude
that the moduli space of vacua is given by all values of $M^{ij}= d^i
\cdot d^j$ subject to $\det M=0$, which breaks the gauge group to
$U(1)$.  However, as we move away from the origin we face the
following problem.  Consider the direction in field space where $M$ is
diagonal and has $N_f-1$ non-zero equal eigenvalues $a$.  For $a \gg
\Lambda^2$ some quarks acquire masses of order $a^{N_f-1}/
\Lambda^{2N_f-3}$ while the massive gauge bosons are much
lighter; their mass is of order $\sqrt a$.  In the energy range
between these two values the gauge group is not broken but the quarks
are not in $SO(N_c)$ representations.  This happens because the
interaction \wdyonici\ is not renormalizable.  Therefore, it cannot be
used for large $a$.  Equivalently, in the limit of large $d$ the gauge
symmetry is broken at a high scale and the gauge interactions are
weakly coupled.  However, the superpotential \wdyonici\ leads to
strong coupling for the massive fields.  Therefore, they cannot be
easily integrated out and the classical analysis is misleading.

Near the origin we can analyze the flat directions by first neglecting
\wdyonici.  Then, the theory is similar to the electric theory and has
several branches.  Its oblique confinement branch is described by the
superpotential \obbra\ in the theory with scale $\tilde {\tilde
\Lambda}$; this $W_{oblique}$ differs from \wdyonici\ by a sign.  Adding
$W_{oblique}$ to \wdyonici\ gives $W=0$.  In this branch of the dyonic
theory we thereby recover the flat directions, given by the space of
$M^{ij}$, exactly as in the electric theory, except that in this theory
it has a strongly coupled description.

Consider perturbing the dyonic theory \wdyonici\ by $W_{tree}=\half
mM^{N_fN_f}$.  Near the origin the dynamics is strongly coupled, as in
the electric theory, and we find the multi-monopole point at strong
coupling.  Away from the origin (for $m \ll \Lambda$) we can integrate
out the massive fields.  Their equations of motion give $d^{\hat
i}\cdot d^{N_f}=d^{N_f}\cdot d^{N_f} =0$ for $\hat i=1\dots N_c-2$,
which generically break $SO(N_c)$ to $SO(2)$.  The massless fields are
$\hat M^{\hat i\hat j}=d^{\hat i}\cdot d^{\hat j}$.  However, in the
region $\det\hat M \approx 16 \Lambda _{N_c,N_c-2}^{2(N_c-2)}$, there
are also light charged fields $d^\pm$ coming from $d^{N_f}$.  The
superpotential in the low energy theory is
\eqn\wdyonlow{W=\half m(1-{\det\hat M\over 16 \Lambda
_{N_c,N_c-2}^{2(N_c-2)}})d^+d^-,}
showing that the charged fields $d^\pm$ are massless at $\det\hat
M=16\Lambda _{N_c,N_c-2}^{2(N_c-2)}$.  The fields $d^\pm$ can be
interpreted as the dyons $E^\pm$ of the low energy $N_f=N_c-2$ theory.
These dyons were found in sect. 3.4 by means of a strong coupling
analysis of the electric theory and in sect. 4.1 by a strong coupling
analysis in the magnetic theory.  Here we find these fields in a weak
coupling analysis of the dyonic theory.  This gives a new interpretation
of the oblique confining superpotential \obbra\ -- it is present in the
tree level Lagrangian of the dyonic theory \wdyonici.

Taking the magnetic dual of the dyonic dual \wdyonici\ gives an $SO(3)$
theory with $N_f$ quarks and the superpotential
\eqn\wmncmid{W={1\over 2\mu}M^{ij}q_i\cdot q_j-{1\over 64\tilde{\tilde
\Lambda}_{N_c, N_c-1}^{2N_c-5}}\det M+{1\over 32\tilde{\tilde
\Lambda}_{N_c, N_c-1}^{2N_c-5}}\det M,}
where the first two terms are as in \wmncmi\ and the last term is the
tree level term \wdyonici\ of the dyonic theory.  This magnetic theory
is the same as the one in \wmncmi ; in particular, using \ddscri\ and
\mgiiisc , the scale and superpotential are the same.
Taking the dyonic dual of the dyonic dual \wdyonici\ shifts the theta
angle by $\pi$ again and gives a superpotential which cancels \wdyonici;
this gives back the original electric theory.  To summarize, the
$SO(N_c)$ theory with $N_c-1$ flavors has three descriptions: the
original electric one, the magnetic $SO(3)$ one discussed in sect. 4.1,
and the dyonic $SO(N_c)$ one with theta angle shifted by $\pi$ and the
superpotential \wdyonici.  Taking duals of duals permutes these three
descriptions.

\centerline{{\bf Acknowledgments}}

We would like to thank T. Banks, R. Leigh, M.R. Plesser, P. Pouliot,
S. Shenker, M. Strassler and E. Witten for useful discussions.  This
work was supported in part by DOE grant \#DE-FG05-90ER40559.

\listrefs
\end